\documentclass[10pt,journal,compsoc]{IEEEtran}
\ifCLASSOPTIONcompsoc
  \usepackage[nocompress]{cite}
\else
  \usepackage{cite}
\fi
\ifCLASSINFOpdf
\else
\fi

\usepackage{microtype}                 
\PassOptionsToPackage{warn}{textcomp}  
\usepackage{textcomp}                  
\usepackage{mathptmx}                  
\usepackage{times}                     
\usepackage{cite}                      
\usepackage{tabu}                      
\usepackage{booktabs}                  

\usepackage{xcolor,colortbl} 

\usepackage{balance}

\usepackage{todonotes}
\usepackage{soul}

\usepackage{nomencl}
\makenomenclature

\def\revcolor{black} 
\newcommand{\rev}[1]{\textcolor{black}{#1}}
\newcommand{\mr}[1]{\textcolor{\revcolor}{#1}}

\begin{document}
%
\title{\rev{Text Entry Performance and Situation Awareness of a Joint Optical See-Through Head-Mounted Display and Smartphone System}}
%
%
%
%


\author{Jens~Grubert,~\IEEEmembership{Member,~IEEE,}
        Lukas~Witzani,
        Alexander~Otte,
        Travis~Gesslein,
        Matthias~Kranz,
        and~Per~Ola~Kristensson

\IEEEcompsocitemizethanks{\IEEEcompsocthanksitem Jens Grubert is and Alexander Otte and Travis Gesslein were with Coburg University of Applied Sciences and Arts, Germany.
\protect\\
E-mail: jens.grubert@hs-coburg.de
\IEEEcompsocthanksitem Lukas Witzani was and Matthias Kranz is with University of Passau, Germany.
\IEEEcompsocthanksitem Per Ola Kristensson is with Cambridge University.}
\thanks{Manuscript received December, 2022.}}

%
%

\markboth{Transactions on Visualization and Computer Graphics}%
{Grubert \MakeLowercase{\textit{et al.}}: Text Entry Performance and Situation Awareness of a Joint Optical See-Through Head-Mounted Display and Smartphone System}
\IEEEtitleabstractindextext{%
\begin{abstract}
Optical see-through head-mounted displays (OST HMDs) are a popular output medium for mobile Augmented Reality (AR) applications. To date, they lack efficient text entry techniques. Smartphones are a major text entry medium in mobile contexts but attentional demands can contribute to accidents while typing on the go. Mobile multi-display ecologies, such as combined OST HMD-smartphone \rev{systems}, promise performance and situation awareness benefits over single-device use. We study the joint performance of text entry on mobile phones with text output on optical see-through head-mounted displays. A series of five experiments with a total of \rev{86} participants indicate that, as of today, the challenges in such a joint interactive system outweigh the potential benefits. 



\end{abstract}

}

\maketitle

\IEEEdisplaynontitleabstractindextext

%
\IEEEpeerreviewmaketitle

\IEEEraisesectionheading{\section{Introduction}\label{sec:introduction}}

\begin{figure*}[tb]
  \centering
  \includegraphics[width=0.66\columnwidth]{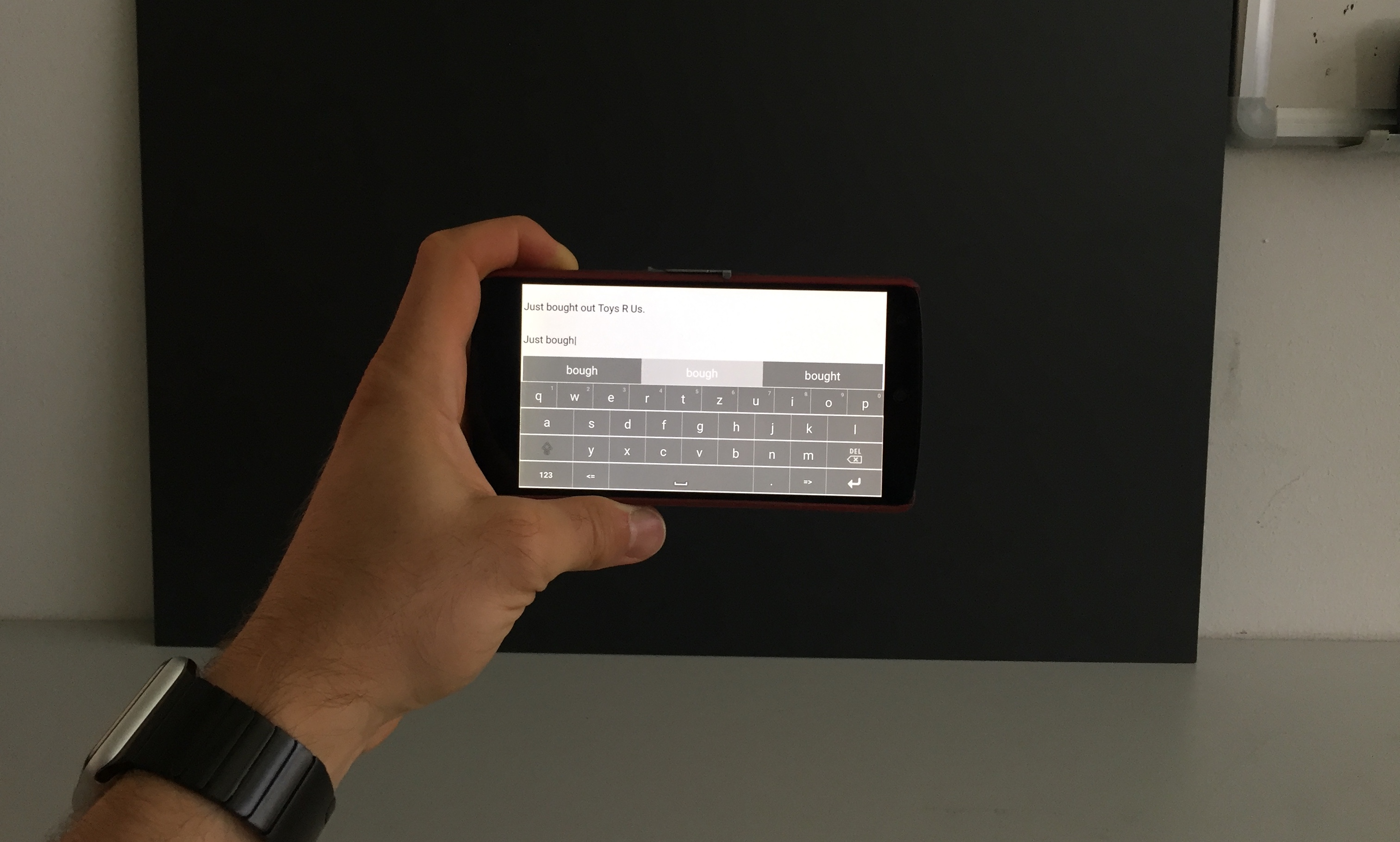}
 	\includegraphics[width=0.66\columnwidth]{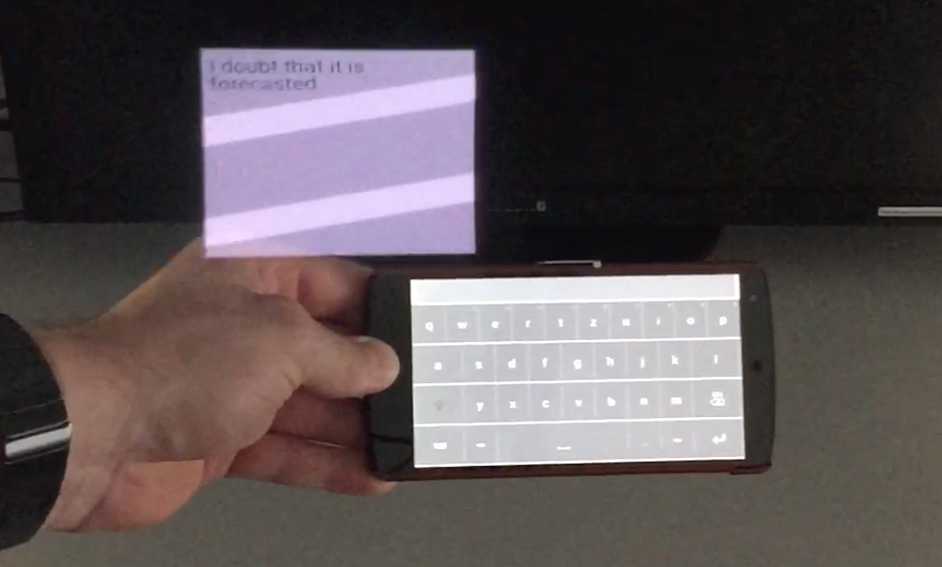}
 	\includegraphics[width=0.66\columnwidth]{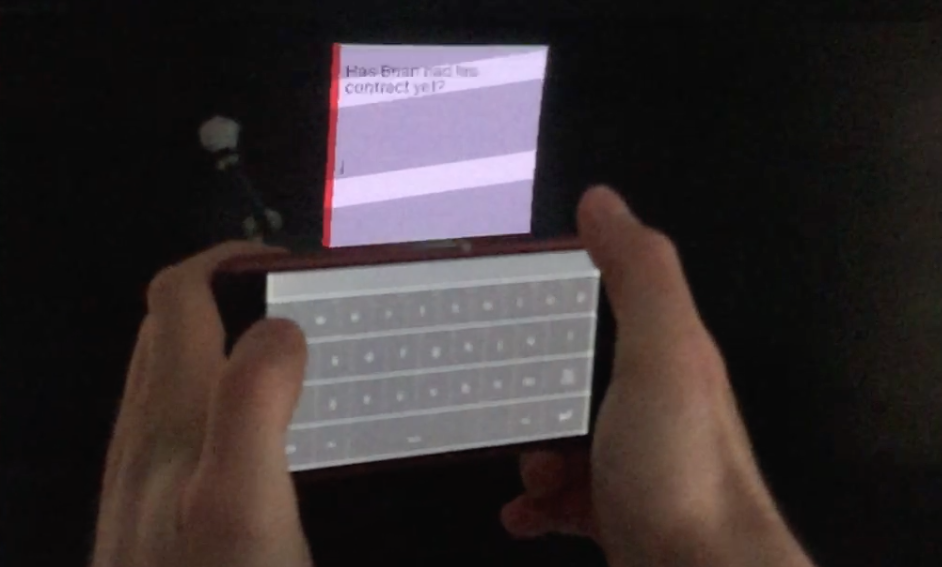}
  \caption{First-person views of the conditions in Experiment 2 on the effects of combined typing on a joint optical see-through head-mounted display (OST HMD) smartphone system. Left: typing on a standard keyboard (condition \textsc{baseline}). Center: typing on a full-screen keyboard displaying the text at a fixed position in the user's field of view (condition \textsc{HUD}). Right: typing on a full-screen keyboard displaying the text spatially registered above the smartphone in an Augmented Reality view (condition \textsc{AR}). The visual artifacts (blur, aberrations, white stripes) are due to capturing the OST HMD screen with a camera.}
 	\label{fig:conditionsinternal}
\end{figure*}

%
%
%
%
\IEEEPARstart{O}{ptical} see-through head-mounted displays (OST HMDs) open up a rich design space for human-computer interaction. OST HMDs can redesign workflows and streamline user experiences in such disparate areas as construction engineering, smart factories and triage systems. In addition, OST HMDs can potentially become the smartphone platform of the future, allowing mobile user experiences to be situated fluidly within physical reality.

However, OST HMDs also \rev{pose} design challenges. One such challenge is \rev{efficient} text entry. Regular text entry using only an OST HMD on a virtual keyboard is difficult as the user's typing activities \rev{have} to be inferred from \rev{typically front-mounted} sensors. The task of inferring \rev{users'} typing in such a setup is challenging as recognition accuracy is limited by sensor range and sampling rate, and the accuracy of the inferred hand skeleton. It is also difficult for users to type in thin air as the typing poses are fatiguing, lack natural physical support, \rev{and do not provide any haptic feedback}. Even smartphone typing, a widely practiced activity, provides \rev{limited} haptic feedback, only informing the user of the fact that the user's finger made contact with the capacitive screen. Such feedback is not present in pure OST HMD typing. \rev{In addition, in safety-critical environments, such as construction engineering, smart factories, and triage systems, it is vital that the text entry system is certain and not subject to the noise and recognition errors that are inherent in mid-air typing solutions.}

As a result \rev{of these concerns}, one commercial OST HMD manufacturer, Microsoft, opted to redesign the typing process for their HoloLens product \rev{such that it involved a specific two-step target acquisition process: 1) acquire a key target on the keyboard by moving a head-tracked cursor to the target; and 2) confirm using a specific click gesture performed with the thumb and index finger. This design provides accuracy for a skilled user at the expense of speed and the need for user learning.} 

As OST HMDs are now beginning to slowly emerge as mainstream user interface technologies, an interesting orthogonal solution space is a joint OST HMD-smartphone system for typing (e.g., \cite{grubert2015multifi})  \rev{that allows users to type on a smartphone while observing the text output on the OST HMD}. Such a \rev{joint} system can potentially provide additional advantages for the user. First, it can provide users with additional privacy as text output is completely hidden from onlookers. Second, it can provide users with better situation awareness\footnote{While there are multiple definitions of situation awareness, we follow Endsley's model \cite{endsley2017toward} but focus on the initial perception of elements in a given situation.} and reduce smartphone-induced pedestrian accidents \cite{nasar2013pedestrian, schwebel2012distraction, yadav4047909systematic}. Third, typing on a smartphone is a relatively fast and well-practiced skill among the existing user population with low fatigue.
\rev{Further, in contrast to alternative mobile text entry devices, such as chording keyboards \cite{lyons2004twiddler}, phones with physical keys, or even body-mounted desktop keyboards \cite{pham2019hawkey}, the smartphone is a ubiquitous interaction device readily available to hundreds of millions of users}

In general, a joint OST HMD-smartphone system opens up possibilities for redistributing input and output spaces in such a way that the advantages of each subsystem are maximized. 
However, such joint OST HMD-smartphone system performance has not been robustly investigated. Grubert et al.~\cite{grubert2015multifi} proposed combining HMDs with smartphones but did not report on performance. To the best of our knowledge, we are the first to quantify the joint performance of text input on mobile phones with text output on OST HMDs. Specifically, we study the effects of non-spatially registered text output (i.e., a heads-up display (HUD) view) and spatially registered text output (AR view) compared to the standard text entry user interface on mobile phones (keyboard and text box on the same capacitive screen). Our central contribution is a series of five investigations that provide a nuanced understanding of the opportunities and challenges inherent with a joint OST HMD-smartphone system. 

\rev{Experiment 1 ($n = 16$) served to validate the fact that larger keys on a smartphone lead to higher entry rates. However, Experiment 2 ($n = 18$) demonstrated that this potential benefit did not materialize for a joint OST HMD-smartphone system for a text entry task in a static setting with participants seated. On the contrary, text entry performance was significantly degraded.}  

\rev{The subsequent three experiments investigated if a positive effect of a joint OST HMD-smartphone system on situation awareness may be observed when using the system in a mobile context. Experiments 3 ($n = 24$) and 4 ($n = 14$) revealed that an OST HMD-smartphone system was not superior compared to a smartphone system in terms of text entry performance or situation awareness. Experiment 5 ($n = 14$) further showed that solely wearing an HMD, even though it is turned off, can already negatively impact users' situation awareness.}

\rev{Taken together, our investigations show that while there are theoretical advantages of joint OST HMD-smartphone systems over using exclusively smartphones for text entry, those advantages do not materialize in current generation systems. }

\section{Related Work}





Text entry is a fundamental user interface task and as a consequence, it is unsurprising that there is a large body of prior research in the area, e.g., 
\cite{kristensson2015next,kristensson2009five}. Mobile text entry has been particularly intensively studied and a standard smartphone QWERTY/QWERTZ/AZERTY keyboard might be suitable since it is portable and has a relatively high text entry rate coupled with an acceptably low error rate \cite{gonzalez2009evaluation, reyal2015performance}.

Many improvements have also been considered for smartphone typing, such as more robust auto-correct algorithms that allow users to type entire sentences before decoding \cite{vertanen2015velocitap} or various combinations of word units \cite{vertanen2018impact}, adding additional support for auto-correcting that incorporates information about the user's gait when walking \cite{goel2012walktype}, and systems that allow a user to self-regulate the certainty of their key presses by pressure regulation \cite{weir2014uncertain}.

\subsection{\rev{Text Entry in AR and VR}}
Prior work on providing text entry for virtual and augmented reality can be split into five categories: 1) special hardware, such as gloves, demanding considerable learning effort and limited performance 
e.g.,~\cite{bowman2002text, jiang2019hifinger, brun2020keycube}; 2) typing in thin air or on a user's palm 
e.g.,~\cite{dudley2018fast, dudley2019performance, wang2021investigating}; 3) utilizing standard physical keyboards 
 \cite{kim2004using, grubert2018text, grubert2018effects}. It was also investigated how to enable interaction in mobile environments using physical keyboards 
 \cite{ otte2019towards, otte2019evaluating,  menzner2019capacitive}; 4) using capacitive touchscreens or controllers 
 \cite{gugenheimer2016facetouch, speicher2018selection, norton2021design};
 and 5) head and gaze pointing-assisted text entry, such as the default text entry method in Microsoft HoloLens and the head-pointed supported gesture-keyboard 
\cite{xu2019ringtext, xu2019pointing}. A recent survey on text entry in VR is presented by Dube et al. \cite{dube2019text}.

\subsection{\rev{Situation Awareness and Smartphone Use}}
Prior research has also investigated situation awareness around smartphone use. Oulasvirta et al.~\cite{oulasvirta2005interaction} discovered that users tended to interact in quick bursts when walking in order to maintain awareness of their surroundings. Several situation awareness support systems have been proposed for mobile users, e.g., \cite{hincapie2013crashalert, wen2015we, foerster2014spareeye, liu2017infrasee}. Wen et al.~\cite{wen2015we} and Liu et al.~\cite{liu2017infrasee} focus on detecting ground obstacles using ultrasound and infrared sensors, Hincapi{\'e}-Ramos et al.~ \cite{hincapie2013crashalert} use a depth camera and Foerster et al.~\cite{foerster2014spareeye} use the smartphone back-camera.  Van dam et al. \cite{van2020effects} indicated negative effects of drivers' passive mobile phone usage on situation awareness while driving. \mr{Woodward and Ruiz \cite{woodward2022analytic} conducted a literature review on situation awareness in AR and presented recommendations on the visual design of AR user interfaces for increasing situation awareness. Complementary to prior work, our work focuses on investigating text entry with a smartphone while wearing an optical see-through head-mounted display.}

\subsection{\rev{Text Legibility and Readability on OST HMDs}}
Further, previous work has investigated text readability on OST HMDs. Gabbard et al. \cite{gabbard2006effects} investigated text readability using different text drawing styles, background textures, and lighting, and later extended their investigations to industrial environments \cite{gattullo2015legibility}. Orlosky et al. \cite{orlosky2013dynamic} proposed a
dynamic text alignment system that actively maintains moving text in the user's field of view, partly inspired by early view management techniques by Blaine et al. \cite{bell2001view}. Lucero et al. \cite{lucero2014notifeye} investigated the presentation of icons while walking, but did not focus on text. Rzayev et al. \cite{rzayev2018reading} investigated different text presentation modes while reading and walking, but did not focus on text entry. \rev{Erickson et al.} \cite{erickson2021extended} \rev{indicated that ``dark mode'' user interfaces could have benefits for presenting UI elements in OST HMDs and that user preferences depend on the lighting conditions in the physical environment. Pavanoto et al. } \cite{pavanatto2021we} \rev{identified the need to increase the font size on virtual monitors in OST HMDs to achieve a comparable reading experience as with physical desktop monitors. Rau et al. } \cite{rau2021immersive} \rev{also indicated that reading on an OST HMD is slower compared to a traditional LCD display.} \mr{Orlosky et al. \cite{orlosky2013dynamic} proposed a text management system that dynamically adapts the text location in order to maximize readability.}

\subsection{\rev{Context and Focus Switching in OST HMDs}}

Related work has investigated the effects of context and focus switching in  OST HMDs. 
\rev{Huckauf et al. investigated the costs of context switching between a monocular OST HMD and a CRT placed at the same focus distance} \cite{huckauf2010perceptual}.
Gabbard et al. \cite{gupta2004empirical, gabbard2018effects, arefin2020impact, arefin2022effect}  examined context switching and differing focal distances between a panel display and a monocular OST HMD (or a haploscope) using text-based visual search task. Both context switching and focal distance switching resulted in significantly reduced performance. Winterbottom et al. \cite{winterbottom2007depth} studied multi-focal AR displays and empirically determined suitable focus distances in the context of a flight simulator. \rev{Eiberger et al. studied the impact of focus switching between an OST HMD and a handheld display} \cite{eiberger2019effects} \rev{and found that both task completion time and error rate increased significantly when solving a visual search task jointly over the OST HMD and smartphone, compared to the OST HMD alone. Similarly, Drouot et al.} \cite{drouot2021effect} \rev{studied the effects of context and focal distance switching using an OST HMD. Their results confirmed a negative impact for focal distance switching but not for context switching.} Gabbard et al. also highlighted the need for multi-focal AR displays \cite{gabbard2014behind} with several research prototypes beginning to emerge \cite{dunn2017wide, akcsit2017near, wilson2018high}. For current single-focus HMDs, Oshima et al. \cite{oshima2016sharpview} and Cook et al. \cite{cook2018user} proposed and evaluated a system for adaptive sharpening HMD display content. \rev{Imamov et al. mimicked context switching between a real-world task and an information display in an immersive VR display and quantified the costs of context switching for different distances} \cite{imamov2020display}. \mr{In a recent survey, Koulieris et al. \cite{koulieris2019near} presented an overview of see-through display technologies, such as varifocal or multiplane displays, that can aid with focus switching.}

\subsection{\rev{Joint OST HMD-Smartphone Interaction}}
To the best of our knowledge, no prior research has investigated the text entry performance of joint OST HMD-smartphone systems in depth. However, joint systems have been studied before (e.g., \cite{grubert2015multifi, wenig2017watchthru, mohr2019trackcap, zhu2020bishare, hubenschmid2021stream, langner2021marvis, knierim2021smartphone}), not focusing on text entry. 
The closest prior works are probably by Wolf et al.~\cite{wolf2018performance}, which studied the performance of pointing, crossing, and steering tasks in a joint OST HMD-smartwatch system but importantly did not investigate text entry and Grubert et al. \cite{grubert2015multifi} who proposed a joint OST HMD-smartphone text entry system but did not report on performance. \rev{Recently, Darbar et al. compared four eyes-free text selection techniques in a joint OST HMD-smartphone system} \cite{darbar2021exploring}. \rev{Their results suggest using the smartphone as basic touchpad outperformed alternative techniques (discrete touch, in-air pointing, raycasting)}. \rev{Besides the use of OST HMDs, research has also explored the joint interaction between smartphones and immersive VR HMDS } \cite{normand2018enlarging, bai2021bringing}.

\section{\rev{Study Overview}}
\rev{In this paper, we investigate the performance and situation awareness aspects of a joint OST HMD-smartphone text entry system.} 





\rev{To this end, Experiment 1 ($n = 16$) quantifies the effects of keyboard size for text entry on a smartphone. The study demonstrates that a full-screen keyboard in landscape mode provides significantly higher text entry rates---around 22 words per minute, an increase of around 15\% compared to a keyboard with standard size.}

\rev{Experiment 2 ($n = 18$) then investigate if this performance gain can be transferred to a joint OST HMD-smartphone system. The main findings of Experiment 2 are that when using an HMD for text output in text entry, performance is significantly reduced compared to a standard smartphone baseline with approximately 10\% lower text entry rates for the HUD condition and approximately 21\% lower text entry rates for the AR condition, in addition to higher error rates. HMD-based text entry was also not preferred by participants.} 

\rev{Taken together experiments 1 and 2 indicate that the potential for faster text entry due to a larger keyboard is diminished by the costs of joint visual information processing across two displays.}


\rev{Experiments 3 ($n = 24$), 4 ($n = 14$), and 5 ($n = 14$) investigate text entry performance in a mobile context. Experiment 3 utilizes a physical obstacle course and Experiment 4 a virtual one in order to focus on potential obstacle collisions. Experiment 3 shows that participants attend the smartphone significantly less when wearing an HMD (41\% of the time) compared to a smartphone-only condition (75\%) or a dual condition (71\%), in which participants could see the text both on the smartphone and the HMD. However, this focus on the HMD induced significantly higher costs in terms of situation awareness (30\% higher cognitive demand) and lower text entry rate (13\%) compared to smartphone-only use.} 

\rev{Experiment 4 indicates an equivalence in object collisions in a simulated walking scenario between a smartphone-only and a joint OST HMD-smartphone not indicating an advantage of wearing an HMD for detecting obstacles while entering text. In addition, wearing an HMD also resulted in higher (but still mild) simulator sickness scores. Experiment 5 replicates the smartphone-only condition and studies the effect of wearing an HMD---without displaying any visual information on it. The experiment indicates equivalence for text entry performance, overall demand, and simulator sickness. However, it also indicates significantly lower attentional supply, in terms of arousal, spare mental capacity, concentration, and division of attention, when wearing an HMD.} \rev{Together, experiments 3, 4, and 5 indicate that, in a mobile condition, combined text entry across a smartphone and an OST HMD does not offer substantial benefits over text entry on a smartphone alone.}
\mr{Unless otherwise indicated, statistical significance tests were carried out using general linear model repeated measures analysis of variance with Holm-Bonferroni adjustments for multiple comparisons at an initial significance level $\alpha = 0.05$. We indicate effect sizes whenever feasible ($\eta^2_p$). We verified that the assumptions underpinning the tests, such as normality and sphericity, were met.}
\rev{
To aid replication and further analysis we make the anonymized experimental data available under https://gitlab.com/mixedrealitylab/ar-text-entry.}





 \section{Experiment 1: Keyboard Size}
\label{ex:keyboardsize}


\rev{One potential benefit of a joint OST HMD-smartphone system is the extended screen space available through the HMD. Prior work \cite{grubert2015multifi} suggested separating text input (on the smartphone) from the output (on the HMD) but did not validate this design. Hence, as a first step, we investigated the effects of different keyboard sizes on a smartphone. While prior work studied the effect of keyboard (key) size on typing performance, e.g., \cite{sears1993investigating, kim2013effects}, we saw it as crucial to quantify the effects at the particular operating point of a smartphone that would be used in a joint HMD-smartphone system. Specifically, based on prior work \cite{grubert2015multifi}, we studied two-handed text entry while the smartphone was held in landscape mode. To this end, we compared a standard-sized onscreen keyboard where a proportion of the space of the display is reserved for a text box, vs. a full-screen onscreen keyboard occupying the entire display. In this experiment, no HMD was used to mitigate the potential effects of focus distance switching, which would occur in an HMD with a fixed focal distance that strongly diverges from the focus plane of a handheld display \cite{eiberger2019effects}.}







\subsection{Method}
The experiment was a within-subjects design with one independent variable: \textsc{KeyboardSize}. The independent variable \textsc{KeyboardSize} had two levels: typing on a normal-sized keyboard (\textsc{StandardKeyboard}), \mr{see Figure \ref{fig:appartusex4keyboard} and Figure \ref{fig:conditionsinternal}, left}, on a smartphone and typing on a larger full-screen keyboard (\textsc{FullscreenKeyboard}) on a smartphone, see Figure \ref{fig:conditionsinternal}, center and right. 

\subsection{Participants}
For this study, we recruited 17 participants from a university campus. One participant had to be excluded due to not being able to write on a smartphone.  All other participants were familiar with smartphone keyboard typing. None had participated in the previous experiments. From the 16 remaining participants (mean age 26.4 years, sd = 7.5, mean height 173.3 cm, sd = 10.4, 6 male, 10 female), Five indicated to write between 0.1-1 h daily using their smartphone, 7 to write between 1-2 h and 2 to write 2-4 h per day. 
Ten participants indicated having visual restrictions which were corrected by contact lenses or glasses during the experiment. 

\subsection{Apparatus and Materials}
\begin{figure}[tb]
 	\centering
 	\includegraphics[width=0.85\columnwidth]{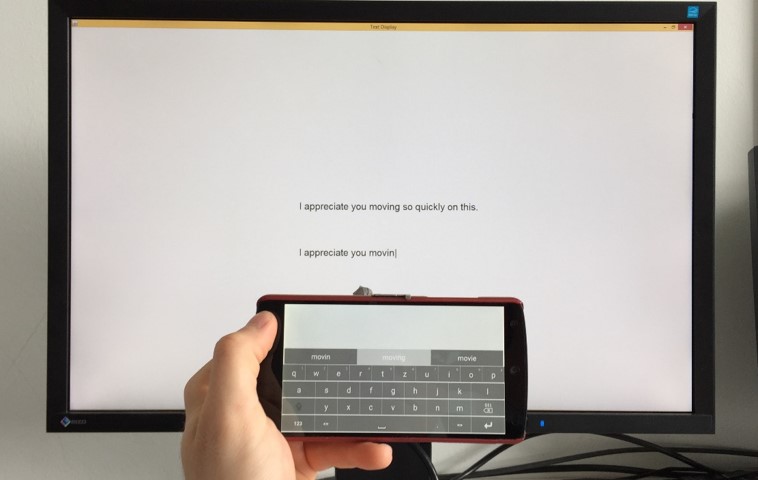} 
 	\caption{The setup used for Experiment 1 on the effect of keyboard size (condition \textsc{StandardKeyboard} is depicted).}
 	\label{fig:appartusex4keyboard}
 \end{figure}
 
Participants were shown stimulus phrases randomly drawn from \rev{a mobile email phrase set \cite{vertanen2011versatile}} on an external monitor (see Figure \ref{fig:appartusex4keyboard}). To ensure comparability between normal and full-screen keyboard, the stimulus text as well as the entered text by the user was shown on an external PC monitor and not on the smartphone directly. An LG Nexus 5 was used as the smartphone. 

\mr{The software keyboard was implemented in two different sizes. The dimensions of the half-screen-sized keyboard were ($w \times h$) 109 $\times$ 31 mm and the dimensions of the keys were 11.0 $\times$ 7.6 mm, see Figure \ref{fig:conditionsinternal}, left. The dimensions of the full-screen keyboard were ($w \times h$) 109 $\times$ 53 mm and the dimensions of the keys were 11 $\times$ 13 mm, see Figure \ref{fig:conditionsinternal}, center.}



\mr{Both keyboards used the same layout. Arrow buttons were added to be able to move the cursor to a different text position during writing. A view showing three suggestions for the currently composed word was shown above the keyboard. The leftmost suggestion showed the currently typed word by the user while the middle and rightmost suggestions showed the most likely and second most likely word respectively. All suggestions could be selected by the user. When a suggestion was selected the keyboard switched the currently composed word into the selected suggestion and moved the cursor to the end of the word. Both keyboards used auto-correction. In the case the user entered a word separator (a space or punctuation character), the keyboard automatically replaced the currently composed word with the most likely word shown as the top middle suggestion. If the user explicitly selected the leftmost suggestion, the auto-correction would not replace the word.}

 

\subsection{Procedure}
Each participant filled out a demographic questionnaire, and was then shown a five-minute video containing information about the study, the participant's task as well as explanations of how the keyboard works. 
The order of the conditions was counterbalanced across all participants. In either condition, participants were shown a series of stimulus sentences. For an individual stimulus sentence, participants were asked to enter it as quickly and as accurately as possible. Participants were allowed to use the backspace key to correct errors.  
Participants typed stimulus sentences for a total of 15 minutes in each condition with a short 30-second break after every 5 minutes of typing. The conditions were separated by a 5-minute break in which participants filled out the NASA TLX questionnaire \cite{hart1988development}. A final questionnaire about the participant's self-assessment was filled out after the second condition. The experiment was carried out in a single 60-minute session structured as a 15-minute introduction and briefing phase, a 40-minute testing phase (15 minutes per condition + five-minute breaks including questionnaires), and five minutes for final questionnaires, an interview, and debriefing.


\subsection{Results}

\begin{table}[t]
    \centering 
    \caption{Descriptive statistics and hypothesis test statistics for error rate and workload for Experiment 1. Grey rows indicate significant differences. CER: Character Error Rate, SK: \textsc{StandardKeyboard}, FK: \textsc{FullscreenKeyboard}, MD: Mental Demand, PD: Physical Demand, TD: Temporal Demand, P: Performance, E: Effort, F: Frustration, O: Overall Demand.}
    \small
    \setlength{\tabcolsep}{5pt}
        
        \begin{tabular}{|c||c|c|c|c|c|}
            \hline 
            & mean (sd) SK & mean (sd) FK & $Z$ & $p$ &  $r$    \\
            \hline 
            CER & $0.008~(0.01)$ & $ 0.007~(0.1) $  & $-0.26$  &  $0.80$ & $-0.05$    \\
            \hline 
            MD & $52.81~(25.43)$ & $ 47.19~(28.75) $  & $1.90$  &  $0.057$ & $0.34$    \\
            \hline 
            \rowcolor{lightgray}
            PD & $48.75~(30.19)$ & $ 36.56~(25.8) $  & $2.30$  &  $0.03$ & $0.41$    \\
            \hline
            TD & $47.50~(15.71)$ & $ 49.38~(20.89) $  & $-0.08$  &  $0.75$ & $-0.01$    \\
            \hline
            \rowcolor{lightgray}
            P & $57.81~(26.20)$ & $ 44.69~(29.07) $  & $1.953$  &  $0.04$ & $0.35$    \\
            \hline
            \rowcolor{lightgray}
             E & $ 60.94~(26.28)$ & $ 51.56~(24.95) $  & $2.50$  &  $0.013$ & $0.44$    \\
            \hline
            \rowcolor{lightgray}
            F & $ 59.06~(22.53) $ & $ 34.06~(22.53) $  & $3.40$  &  $>0.001$ & $0.60$    \\
            \hline
            \rowcolor{lightgray}
            O & $ 54.48~(21.32) $ & $ 43.01~(21.49) $  & $2.95$  &  $0.003$ & $0.52$    \\
            \hline 
        \end{tabular}
     \label{tab:results_e1_cer_tlx}
\end{table}


\textbf{Text Entry Rate and Error Rate:} Entry rate was measured in words-per-minute (wpm), with a word defined as five consecutive characters, including spaces. The time frame used for calculating the entry rate for each sentence started at the appearance of the stimulus text and ended when hitting the enter button.

For \textsc{StandardKeyboard}, the mean entry rate was 19.51 wpm (sd = 4.4). For \textsc{FullscreenKeyboard}, the mean entry rate was 22.37 wpm (sd = 4.45). A paired two-tailed t-test showed that the entry rate difference between \textsc{StandardKeyboard} and \textsc{FullscreenKeyboard} was statistically significant ($t(15) = 3.5862$, $p = 0.0027$) with an effect size of $Cohen's~d_z = 0.93 $ (calculated after \cite{lakens2013calculating}).

Error rate was measured as character error rate (CER). CER is the minimum number of character-level insertion, deletion, and substitution operations required to transform the response text into the stimulus text, divided by the number of characters in the stimulus text. Results for character error rate can be seen in Table \ref{tab:results_e1_cer_tlx}. No significant statistical difference was detected. \rev{In other words, \textsc{FullscreenKeyboard} resulted in significant higher text entry rates compared to \textsc{StandardKeyboard}.}

\textbf{Workload: } Scores for the workload of both conditions measured by the unweighted NASA TLX questionnaire are depicted in \rev{Table  \ref{tab:results_e1_cer_tlx}}. Wilcoxon signed-rank tests did reveal significant differences for physical demand, performance, effort, frustration, and overall demand, but not for mental demand and temporal demand. \rev{In other words, \textsc{FullscreenKeyboard} resulted in significant lower workload compared to \textsc{StandardKeyboard}.}



\textbf{Preferences and Open Comments: } Participants were asked to rank the conditions from least preferred to most preferred. Ten out of the 16 participants preferred the bigger keyboard. \rev{A binomial test did not indicate a significant difference ($p = 0.454$)}.  Ten participants felt increased typing speed with the full-screen keyboard. 15 participants estimated to mistype more often when using the small keyboard.
Qualitative feedback also revealed that participants found the big keyboard to be easier to use and its buttons easier to hit. They felt to have more precision and needing to correct mistakes less often. Reasons for preferring the small keyboard were either the habit of using this size for a keyboard or the smaller distance the finger had to travel while typing. 





\section{Experiment 2: Effects of Combined Typing on a Joint OST HMD-Smartphone System}

Experiment 2 investigated the effects of combined typing on a smartphone and an OST HMD. We compared QWERTZ keyboard text entry on a mobile phone \mr{(Figure \ref{fig:conditionsinternal}, left)} with text entry on a joint OST HMD-smartphone system in two configurations. In the first OST HMD-smartphone configuration, as proposed by Grubert et al. \cite{grubert2015multifi}, we used a full-screen keyboard on the smartphone for text entry combined with a spatially registered AR view on an OST HMD \mr{(Figure \ref{fig:conditionsinternal}, right)}. In the second OST HMD-smartphone configuration, we investigated a non-spatially registered view (HUD), see Figure \ref{fig:conditionsinternal}, \mr{center}. 
\mr{Hence, text input was always conducted on the smartphone, but text output appeared either on the smartphone or on the OST-HMD (spatially registered or fixed).}
\rev{In addition, to Experiment 1, we also added NASA TLX \cite{hart1988development}, in order to receive workload indications, the Simulator Sickness Questionnaire (SSQ) \cite{kennedy1993simulator}, in order to investigate if the spatially registered text might lead to higher simulator sickness, and the Flow-Short-Scale questionnaire (FSS) \cite{rheinberg2002measuring}, in order to investigate if the HMD condition would lead to a higher flow.} 



\subsection{Method}
The experiment was a within-subjects design with one independent variable \textsc{interface}.
The independent variable \textsc{interface} had three levels: \textsc{Baseline}, \textsc{HUD}, and \textsc{AR}. In the \textsc{Baseline} condition, text entry was done using the half-screen-sized mobile keyboard with the text shown on the smartphone itself. In the \textsc{HUD} condition, text entry was done using the full-screen keyboard and text output was on the OST HMD with a HUD-like display, i.e. with screen-aligned text that was not spatially registered. Third, in the \textsc{AR} condition, text entry was also done using the full-screen keyboard, but text output was spatially registered to the smartphone as proposed by Grubert et al. \cite{grubert2015multifi}. In this mode, the user sees the display through the glasses as a virtual extension to the smartphone screen, i.e. only if the head points toward the smartphone. The order of the starting conditions was balanced across participants. The task was to write as quickly and as accurately as possible. 

\subsection{Participants}
For this study, we recruited 18 participants (mean age 22.9 years, sd = 1.54, mean height 172.7 cm, sd = 7.79, 5 male, 13 female) from a university campus.
Eleven participants indicated having visual restrictions: ten used contact lenses to correct these and one participant used her glasses during the experiment. 
The interpupillary distance was measured with a ruler (mean = 63.06 cm, sd = 3.24) to adjust the virtual camera positions for the \textsc{AR} condition.
All participants were familiar with smartphone keyboard typing. Fourteen participants never used an HMD before, two once, and two rarely. 

\subsection{Apparatus and Materials}

Stimulus sentences were drawn from a mobile email phrase set \cite{vertanen2011versatile}. Participants were shown stimulus phrases randomly drawn from the set. An LG Nexus 5 was used as the smartphone. \mr{We used the same software keyboard as in Experiment 1.}

An Epson Moverio BT-300 (23\textdegree diagonal field of view, 1280 $\times$ 720 pixel resolution) was used as the OST HMD. Participants were sitting on a chair and could rest their arms on a table if they wanted to. 
\mr{The font size was set empirically to approximately 0.9\textdegree on the HMD, such that the text was still legible (and matched on the smartphone accordingly).}
An Optitrack Flex 13 outside-in tracking system was used for spatial tracking of the smartphone and HMD for spatial tracking in the \textsc{AR} condition. For this purpose, the smartphone and the OST HMD were equipped with retroreflective markers. \mr{We used a fixed optical see-through calibration for all users (Single Point Active Alignment, SPAAM \cite{tuceryan2002single}) but ensured that all participants had a correct spatially registered view in the \textsc{AR} condition.} 

\subsection{Procedure}

Each participant filled out a demographic questionnaire. Then, each participant was shown a five-minute video containing information about the study, the study task, as well as explanations on how the keyboard worked. 
The interpupillary distance was measured to ensure correct stereo rendering for the \textsc{AR} condition. 
Participants typed stimulus sentences for a total of 15 minutes in each condition with a short 30-second break after every 5 minutes of typing. In the 5-minute break between conditions, participants filled out the unweighted NASA TLX \cite{hart1988development}, the Simulator Sickness Questionnaire (SSQ) \cite{kennedy1993simulator} and the Flow-Short-Scale questionnaire (FSS) \cite{rheinberg2002measuring}. Participants also filled out a  final self-assessment questionnaire after the last condition, followed by a semi-structured interview. The conditions were balanced across participants. In either condition, participants were shown a series of stimulus sentences and asked to type them as quickly and as accurately as possible. 
Participants were allowed to use the backspace key to correct errors. The experiment was carried out in a single 85-minute session structured as a 20-minute introduction and briefing phase, a 60-minute testing phase (15 minutes per condition + five-minute breaks including questionnaires), and 5 minutes for filling out the final questionnaire, the interview, and debriefing.

\subsection{Results}

\begin{table}[t!]
    \centering 
    \caption{Descriptive statistics and hypothesis test statistics for \mr{text entry rate} and error rate for Experiment 2. Grey rows indicate significant differences. B: \textsc{Baseline}, WPM: Words Per Minute. CER: Character Error Rate.}
    \scriptsize
    \setlength{\tabcolsep}{5pt}
        
        \begin{tabular}{|c||c|c|c|c|c|c|}
            \hline 
            & mean (sd) B & mean (sd) HUD  & mean (sd) AR & $F_{2,16}$  & $p$ &  $\eta^2_p$    \\
            \hline 
            \rowcolor{lightgray}
            \mr{WPM} & $22.73~(4.02)$ & $20.57~(4.65) $  & $17.95~(4.68)$  &  $12.68$ & $0.001$ & $0.61$   \\
            \hline 
            \rowcolor{lightgray}
            CER & $0.007~(0.006)$ & $ 0.021~(0.016) $  & $0.097~(0.07)$  &  $9.30$ & $0.002$ & $0.54$    \\
            \hline 

        \end{tabular}
     \label{tab:results_e2_tct_cer}
\end{table}

\textbf{Text Entry Rate and Error Rate: } Descriptive statistics and results of RM-ANOVA omnibus tests for entry rate and character error rate are depicted in Table \ref{tab:results_e2_tct_cer}.
For text entry rate, Holm-Bonferroni adjusted post-hoc testing revealed that there were significant differences between \textsc{Baseline} and \textsc{HUD} ($p = 0.006$) and \textsc{AR} ($p < 0.001$), as well as between \textsc{HUD} and \textsc{AR} ($p = 0.003$).
For character error rate, Holm-Bonferroni adjusted post-hoc testing revealed that there were significant differences between \textsc{Baseline} and \textsc{HUD} ($p = 0.002$) and \textsc{AR} ($p = 0.002$) as well as between \textsc{HUD} and \textsc{AR} ($p = 0.004$).
\rev{In summary, \textsc{HUD} resulted in a 10\% lower text entry rate and three times higher character error rate (albeit on a low absolute level) compared to \textsc{Baseline}. \textsc{AR} resulted in a 21\% lower text entry rate and a 14 times higher character error rate compared to \textsc{Baseline}.}

\begin{table}[t!]
    \centering 
    \caption{Descriptive statistics and hypothesis test statistics for workload, \mr{simulator sickness, and flow} for Experiment 2. Grey rows indicate significant differences. B: \textsc{Baseline}, MD: Mental Demand, PD: Physical Demand, TD: Temporal Demand, P: Performance, E: Effort, F: Frustration, O: Overall Demand. NA: Nausea, OM: Oculo-Motor, DI: Disorientation, TS: Total Severity, FF: Flow, FA: Anxiety subscale of FSS, FC: Challenge subscale of FSS.}
    \scriptsize
    \setlength{\tabcolsep}{5pt}
        
        \begin{tabular}{|c||c|c|c|c|c|c|}
            \hline 
            & mean (sd) B & mean (sd) HUD  & mean (sd) AR & $\chi^2(2)$  & $p$    \\
            \hline 
            \rowcolor{lightgray}
            MD & $35.28~(20.25)$ & $42.5~(20.88) $  & $58.61~(21.95)$  &  $24.80$ & $<0.001$  \\
            \hline 
            \rowcolor{lightgray}
            PD & $27.50~(19.95)$ & $33.61~(20.42) $  & $49.72~(27.63)$  &  $24.49$ & $<0.001$    \\
            \hline 
            TD & $ 48.33~(23.01)$ & $ 43.89~(17.87) $  & $48.06~(23.34)$  &  $0.49$ & $0.79$    \\
            \hline 
            \rowcolor{lightgray}
            P & $ 35.56~(13.60)$ & $ 41.11~(15.10) $  & $51.11~(22.12)$  &  $9.94$ & $0.007$    \\
            \hline 
            \rowcolor{lightgray}
            E & $ 44.17~(21.71)$ & $ 50.83~(17.34) $  & $ 63.61~(19.46)$  &  $8.85$ & $0.012$    \\
            \hline 
            \rowcolor{lightgray}
            F & $ 36.94~(16.37)$ & $ 41.39~(19.61) $  & $ 60.83~(20.45)$  &  $28.90$ & $<0.001$    \\
            \hline 
            \rowcolor{lightgray}
            O & $37.96~(13.66)$ & $ 42.22~(13.66) $  & $ 55.32~(16.20)$  &  $23.01$ & $<0.001$    \\
            \hline 
            \hline
            \rowcolor{lightgray}
            NA & $44.3~(44.4)$ & $ 47.4~(37.9) $  & $71.5~(50.8)$  &  $12.03$ & $0.002$    \\
            \hline
            \rowcolor{lightgray}
            OM & $21.5~(31.8)$ & $ 21.9~(28.5) $  & $28.6~(33.4)$  &  $8.33$ & $0.016$    \\
            \hline
            \rowcolor{lightgray}
            DI & $43.3~(28.3)$ & $ 46.4~(43.0) $  & $82.7~(73.6)$  &  $10.34$ & $0.006$    \\
            \hline
            \rowcolor{lightgray}
            TS & $44.3~(44.4)$ & $ 47.4~(37.9) $  & $71.5~(50.8)$  &  $9.77$ & $0.008$    \\
            \hline
            \hline
            FF & $4.56~(1.14)$ & $ 4.54~(1.05) $  & $3.99~(1.13)$  &  $5.51$ & $0.063$    \\
            \hline
            FA & $3.13~(1.65)$ & $ 2.87~(1.53) $  & $ 2.94~(1.60)$  &  $2.51$ & $0.285$    \\
            \rowcolor{lightgray}
            FC & $ 3.50~(0.86)$ & $ 3.67~(0.77) $  & $ 4.61~(1.15)$  &  $16.69$ & $<0.001$    \\

\end{tabular}
     \label{tab:results_e2_tct_cer}
\end{table}

 
\textbf{NASA TLX: } 
The scores as well as the statistics of Friedman omnibus tests for workload as measured by Nasa TLX are depicted in Table \ref{tab:results_e2_tct_cer} and descriptive statistics in Table \ref{tab:results_e2_tct_cer}. 
Post-hoc analysis with Wilcoxon signed-rank tests and Holm-Bonferroni correction revealed that there were significant differences for mental demand between \textsc{AR} and \textsc{baseline} ($p < 0.001$), \textsc{AR} and \textsc{HUD} ($p < 0.001$)
, for physical demand between all conditions ($p < 0.016$), for performance between \textsc{AR} and \textsc{HUD} ($p = 0.005$), for effort, frustration and overall demand between \textsc{AR} and \textsc{baseline} ($p < 0.001$). No other significant differences were indicated. \rev{
In other words, \textsc{AR} led to significantly higher mental and physical demand compared to both \textsc{Baseline} and \textsc{HUD}, lower performance compared to \textsc{HUD}, as well as higher effort, frustration, and overall demand compared to \textsc{Baseline}.
}

\textbf{Simulator Sickness: } 


\begin{table}
\caption{Reference SSQ score ranges for none to severe symptoms.}
\label{tab:ssqref}
\begin{center}
    \small
	\begin{tabular}{ |c|c|c|c|c| }
		\hline
		Level & Nausea & Oculo-motor & Disorientation & Total \\
		\hline
		\hline
		none & 0  & 0 & 0 & 0  \\
		\hline
		slight & 66.8 & 53.1 & 97.4 & 78.5 \\
		\hline
		moderate & 133.6 & 106.1 & 194.9 & 157.1\\
		\hline
		severe & 200.3 & 159.2 & 292.3 & 235.6 \\
		\hline
	\end{tabular}
 \vspace{-0.5cm}
\end{center}

\end{table}

The scores for the SSQ scales and for a Friedman omnibus test are depicted in Table \ref{tab:results_e2_tct_cer}. 
For reference purposes, the possible score ranges are depicted in Table \ref{tab:ssqref}. 
Post-hoc analysis with Wilcoxon signed-rank tests and Holm-Bonferroni correction revealed that there were significant differences for oculo-motor between \textsc{AR} and \textsc{baseline} ($p = 0.009$), \textsc{AR} and \textsc{HUD} ($p = 0.009$), for nausea, between \textsc{AR} and \textsc{baseline} ($p = 0.007$) and \textsc{AR} and \textsc{HUD} ($p = 0.006$), for disorientation, between \textsc{AR} and \textsc{baseline} ($p = 0.006$) and \textsc{AR} and \textsc{HUD} ($p = 0.003$), and, for total severity, again between \textsc{AR} and \textsc{baseline} ($p = 0.004$) and \textsc{AR} and \textsc{HUD} ($p = 0.006$). No other significant differences were indicated.
\rev{ In other words, \textsc{AR} led to significantly higher SSQ scores in all dimensions compared to both \textsc{Baseline} and \textsc{HUD}. Symptoms can be regarded as slight (total severity, oculo-motor, disorientation for all conditions) to moderate (nausea for \textsc{AR}).
}



\textbf{Flow: } 
The scores for the flow scales and for a Friedman omnibus test are depicted in Table \ref{tab:results_e2_tct_cer}.
Post-hoc analysis with Wilcoxon signed-rank tests and Holm-Bonferroni correction revealed that there were significant differences for challenge between \textsc{AR} and \textsc{baseline} ($p = 0.001$) as well as between \textsc{AR} and \textsc{HUD} ($p = 0.004$). No other significant differences were indicated.
\rev{In other words, \textsc{AR} led to a significantly lower flow score compared to both  \textsc{HUD} and \textsc{Baseline}.}


\textbf{Preferences and Open Comments: }
When asked to rank all three conditions, 
\rev{13} out of 18 participants preferred \textsc{Baseline}, \rev{four} \textsc{HUD} and \rev{one} \textsc{AR}.
Five participants preferred the \textsc{baseline} condition due to habits. 

\rev{Friedman's tests revealed statistically significant differences on ranks} ($\chi^2(2) = 17.4$, $p < 0.001$). \rev{Post-hoc analysis revealed that there were significant differences between all conditions} ($p < 0.02$).

Also, four didn't enjoy wearing the OST HMD in general. Five participants explicitly preferred \textsc{HUD} over \textsc{AR} because of the display being always visible. They liked the freedom of head movements and disliked the restrictions that \textsc{AR} would force on them. Two felt the display to be sharper in \textsc{HUD} than in \textsc{AR} mode. Positive comments on \textsc{AR} included comments about the overlay being in the same plane as the smartphone's screen. Two participants liked this mode more because it felt easier for them. \textsc{AR} was indicated to be the most exhausting of all three types of display. People disliked the fact that they would need to have the smartphone at one exact position in relation to the head to be able to read the text. 

When asked if the participants would use OST HMDs for privacy reasons during text entry on smartphones, results of answering a 10-item Likert scale (1 not at all - 10 at all times) resulted in a mean of 3.76 (sd = 1.97). Reasons for this varied. Seven people indicated not to see a need for privacy, and, if so, there would exist other ways to achieve privacy e.g., covering the smartphone with one's body. Six people described the OST HMD for this use as too inconvenient. Two people wouldn't feel comfortable in public space using the HMD and three described the use as exhausting. However, five people could see use cases where OST HMDs could be useful. Mainly, they saw the use in occupational environments where privacy and secrecy have more important roles than during leisure activities.

\begin{figure}[tb]
	\centering
	 \includegraphics[width=1\columnwidth]{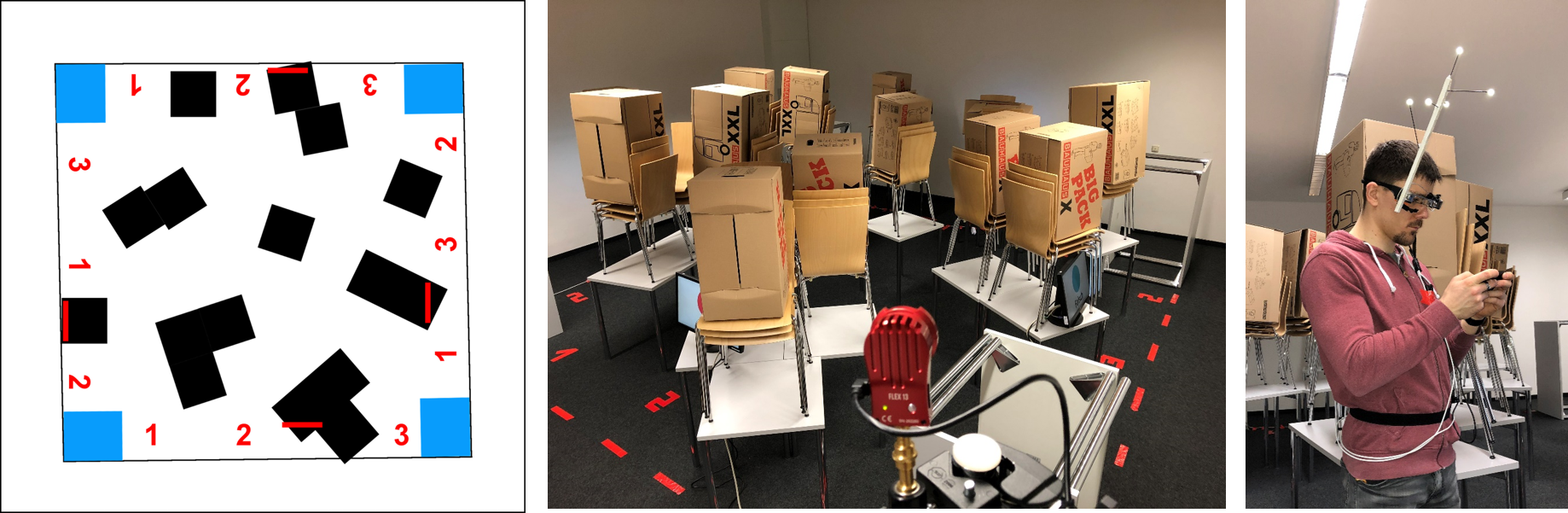}
	\caption{Apparatus for Experiment 3. Left: schematic top-down view on 7 $\times$ 7 m obstacle course (black: tables, red numbers: entries into the course, red line: instruction monitor blue: boundary tables). Center: Obstacle course. Right: User with HMD and smartphone.}	\label{fig:apparatusmaze}
\end{figure}

\subsection{Discussion}
Experiment 2 indicated, that wearing an HMD for text output resulted in significantly lower text entry performance compared to a standard smartphone. 
 Further, the spatially registered \textsc{AR} view performed worse than the \textsc{HUD} condition. 
The lower text entry performance is accompanied by significantly higher workload scores for \textsc{AR} compared both to \textsc{baseline} and \textsc{HUD}, significantly higher nausea and oculo-motor problems of \textsc{AR} compared to  \textsc{baseline} and a higher challenge level for \textsc{AR} compared both to \textsc{baseline} and \textsc{HUD}.

A potential factor for this outcome is the low vertical field-of-view (ca. 11\textdegree), which forced participants to tilt their heads in order to see the virtual text above the smartphone in the \textsc{AR} condition. Further, the vergence-accommodation conflict potentially influenced participants' performance and well-being. 
In summary, \rev{the joint OST HMD-smartphone system did not provide} performance benefits compared to standard smartphone typing. We conjecture this result fundamentally depends on two different factors: 1) the interaction context; and 2) the parameters of the subsystems in the joint OST HMD-smartphone system. 



\section{Experiments 3, 4 and 5: Text Entry while Walking}

We hypothesized that the static context in \mr{Experiment 2} did not provide an opportunity for the joint HMD-smartphone system to provide any detectable advantages for users. \rev{Specifically, while users could potentially recognize their physical surroundings due to looking heads-up at the OST HMD instead of looking heads-down on a smartphone, this potential benefit could not be validated in the static context of Experiment 2}. 

To better understand this factor, we conducted \rev{three} experiments focusing on user behavior while walking. In Experiment 3, we exposed participants to a physical obstacle course. In Experiment 4, we replaced the physical obstacle course with a virtual one to be able to investigate the effect of obstacle collision risks without endangering participants' health. \rev{Finally, in Experiment 5, we studied if wearing an HMD alone, without any information displayed, would have potential negative impacts on user behaviour. For these experiments we included additional measures --- the number of collisions, gaze, and the Situational Awareness Rating Technique (SART) questionnaire \cite{taylor2017situational}, in order to investigate situation awareness in a mobile context.}


\subsection{Experiment 3: Obstacle Course}

Experiment 3 focused on text entry while walking. To this end, we employed a physical obstacle course, in which participants were asked to walk and type simultaneously.


\subsubsection{Method}

\begin{figure}[tb]
	\centering
	 \includegraphics[width=1\columnwidth]{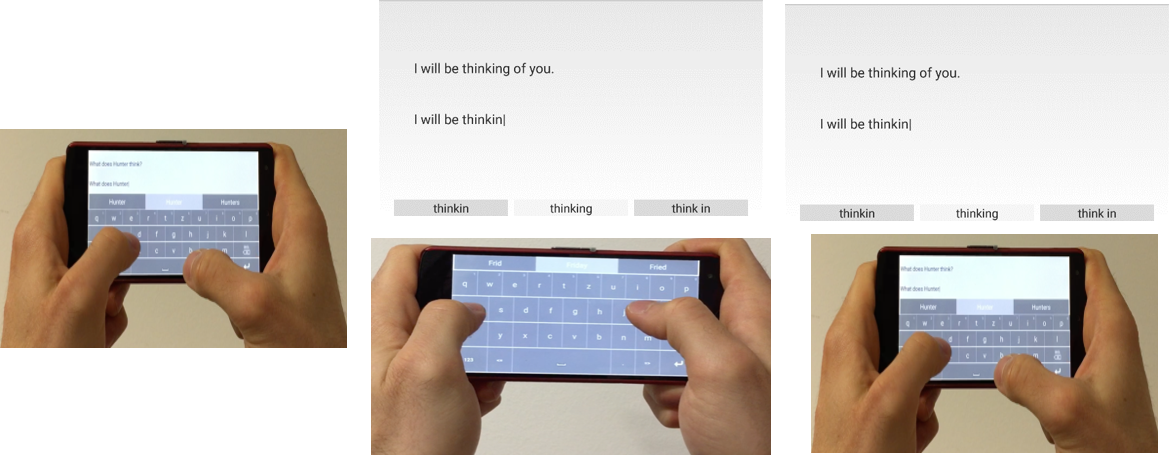}
	\caption{Conditions for Experiment 3, obstacle course. Left: View on smartphone in condition \textsc{baseline}. Center: view on smartphone (bottom) and HMD view (top, as screenshot) for condition \textsc{HUD}, view on smartphone (bottom). Right: HMD view (top, again as screenshot) in condition \textsc{DUAL}. Please note, that the HMD views are depicted as screenshots but were visible to users on the see-through HMD (i.e. not opaque).}	\label{fig:conditionsmaze}
\end{figure}

The experiment was a within-subjects design with one independent variable \textsc{interface}.
The independent variable \textsc{interface} had three levels: \textsc{Baseline}, \textsc{HUD} and \textsc{Dual}. In the \textsc{Baseline} condition, text entry was done using the standard-sized mobile keyboard that was used in the previous experiments. In the \textsc{HUD} condition, as previously investigated in Experiment 2, the full-screen keyboard was used for text input and the HMD for output. As a progression from Experiment 2, we now used an Android view in the HMD to improve text clarity, see Figure \ref{fig:conditionsmaze}. In the \textsc{Dual} condition, text entry was carried out with a standard keyboard on the smartphone, and text output was shown both on the HMD (as in condition \textsc{HUD}) and on the smartphone (as in condition \textsc{baseline}). This was done to investigate user behavior when users are not forced to look at the HMD. The order of the starting conditions was balanced across participants. The task was to write as quickly and as accurately as possible while walking through a physical obstacle course at a self-paced speed. 

\subsubsection{Participants}
In this study, 24 volunteers participated (mean age 24.75 years, sd = 3.14, mean height 175.58 cm, sd = 9.89, 13 male, 11 female). None had participated in the previous experiment. Eight participants indicated to have visual restrictions. Six used contact lenses to correct these. No participant used additional glasses during the experiment. All participants were familiar with typing on a smartphone. Sixteen participants never used an HMD before, 6 once, and two rarely. Data for gaze and text entry metrics from one participant needed to be excluded due to logging errors.

\subsubsection{Apparatus and Materials}
The apparatus is shown in Figure \ref{fig:apparatusmaze}. A physical maze with dimension 7 $\times$ 7~m was used (see Figure \ref{fig:apparatusmaze}, left and middle). On each side, a monitor displayed which entry participants should take into the maze. The participants were then instructed to go either to the monitor on the opposite side of the maze or to a monitor left or right to the current one outside of the maze. \rev{This was done to generate a variety of possible paths through the maze.} We used an LG Nexus 5 as the smartphone and an Epson Moverio BT-300 as the OST HMD. The OST HMD was equipped with retro-reflective markers on a rod to be able to record participants' movements through the maze. \mr{We ensured that the rod had no substantial effect on the comfort of wearing the headset. Users also did not report on any such issues.} Also, the HMD was equipped with a mobile Pupil Labs eye-tracker (connected to another smartphone for recording) \mr{with an accuracy of 0.6\textdegree}. \rev{The text size for the displayed text output was the same across conditions.}

\subsubsection{Procedure}

Before each condition, we calibrated the eye tracker. Participants typed stimulus sentences from the same sentence set as in Experiment 3 for a total of 10 minutes in each condition. \rev{While typing, they walked through the maze and followed instructions on which way to take through the maze. While the instructions, which way to follow, were randomized, it was possible that participants took a specific route multiple times.} There were 7-minute breaks between conditions. During the breaks, participants filled out the same questionnaires as in Experiment 2 (TLX, SSQ, FSF) and, in addition, the SART questionnaire. Towards the end, participants filled out a final self-assessment questionnaire, followed by a semi-structured interview. 
In either condition, participants were shown a series of stimulus sentences \rev{(drawn from the same phrase set as in the other experiments)}. 
\rev{The order of conditions was counterbalanced}. The experiment was carried out in a single 90-minute session structured as a 20-minute introduction and briefing phase
, a 65-minute testing phase (five minutes for calibration + 10 minutes per condition + seven-minute breaks including questionnaires), and 5 minutes for the final questionnaire, interview, and debriefing.

\subsubsection{Results}
\textbf{Text Entry Rate and Error Rate: }

\begin{table}[t!]
    \centering 
    \caption{Descriptive statistics and hypothesis test statistics for \mr{text entry rate and} error rate for Experiment 3. Grey rows indicate significant differences. B: \textsc{Baseline},  H: \textsc{HUD}, D: \textsc{Dual}, WPM: Words Per Minute. CER: Character Error Rate.}
    \scriptsize
    \setlength{\tabcolsep}{5pt}
    \begin{tabular}{|c||c|c|c|c|c|c|}
            \hline 
            & mean (sd) B & mean (sd) H & mean (sd) D & $F_{2,21}$  & $p$ &  $\eta^2_p$    \\
            \hline 
            \rowcolor{lightgray}
            WPM & $18.70~(3.64)$ & $16.24~(3.62)$  & $ 18.05~(4.31)$  &  $ 11.94 $ & $<.001$ & $0.53$   \\
            \hline 
            \rowcolor{lightgray}
            CER & $0.087~(0.082)$ & $ 0.018~(0.015) $  & $ 0.011~(0.010) $  &  $ 5.89$ & $.009$ & $0.53$    \\
            \hline 
        \end{tabular}
     \label{tab:results_e3_tct_cer}
\end{table}  

Descriptive statistics and results of RM-ANOVA omnibus tests for entry rate and character error rate are depicted in Table \ref{tab:results_e3_tct_cer}.
For, text entry rate, Holm-Bonferroni adjusted post-hoc testing revealed that there were significant differences between \textsc{Baseline} and \textsc{HUD} ($p < 0.001$) as well as between \textsc{Dual} and \textsc{HUD} ($p = 0.004$), but not \textsc{Baseline} and \textsc{Dual} ($p = 0.812$).
For character error rate, Holm-Bonferroni adjusted post-hoc testing revealed that there was a significant difference between \textsc{Baseline} and \textsc{HUD} ($p = 0.006$). No other significant differences were detected.
\rev{In other words, \textsc{HUD} led to a significantly lower text entry rate compared to both \textsc{Baseline} and \textsc{Dual}, and to a significantly higher error rate compared to \textsc{Baseline}.}

\textbf{Eye Gaze: } We investigated the duration participants spent looking at the smartphone screen. \mr{To this end, for each frame, we checked if the reported normalized 2D gaze pointer would fall into the bounding box of the smartphone screen.} Data from one participant had to be excluded due to non-working eye-tracking. The mean relative duration (i,e. duration spent looking at the smartphone relative to the duration of the condition) for \textsc{baseline} was 74.53\% (sd = 22.87), for \textsc{HUD} 41.09\% (sd = 17.48 ) and for \textsc{dual} 70.32\% (sd = 19.93), see also Table \ref{tab:results_e3_gaze}. An omnibus test revealed that the difference in gaze duration was statistically significant ($F_{2,20} = 46.244$, $\eta^2_p = 0.822$, $p < 0.001$). Holm-Bonferroni adjusted post-hoc testing revealed that there were significant differences between \textsc{HUD} and \textsc{Baseline} ($p < 0.001$), \textsc{HUD} and \textsc{Dual} ($p < 0.001$), but not between \textsc{Baseline} and \textsc{Dual} ($p = 0.70$). \rev{In other words, participants looked at smartphone content in the \textsc{HUD} condition significantly less compared to both \textsc{Baseline} and \textsc{Dual}.}  


\begin{table}[!t]
    \centering 
    \caption{Relative gaze duration on smartphone for Experiment 3, obstacle course.}
    \setlength{\tabcolsep}{5pt}
    \begin{tabular}{|c||c|c|c|}
            \hline 
            & DUAL B & HUD & baseline \\
            \hline 
            mean & 0.67 B & 0.41 & 0.75 \\
            \hline 
            sd & 0.24 B & 0.17 & 0.22 \\           
            \hline 
        \end{tabular}
     \label{tab:results_e3_gaze}
\end{table}


\textbf{Workload:} The mean mental demand score for workload as measured by Nasa TLX was 54.58 (sd = 19.83) for \textsc{baseline}, 57.29 (sd = 20.95) for \textsc{HUD} and 66.04 (sd = 23.077) for \textsc{Dual}. The mean overall score for workload as measured by Nasa TLX was 47.22 (sd = 15.17) for \textsc{baseline}, 48.85 (sd = 16.19) for \textsc{HUD}, and 53.06 (sd = 16.33) for \textsc{Dual}.  Friedman's tests revealed statistically significant differences for mental demand ($\chi^2(2) = 12.602$, $p = 0.002$) and overall demand ($\chi^2(2) = 6.796$, $p = 0.033$), but not for temporal demand. 
Post-hoc analysis with Wilcoxon signed-rank tests and Holm-Bonferroni correction revealed that there were significant differences for mental demand between \textsc{DUAL} and \textsc{baseline} ($p = 0.002$) as well as for \textsc{DUAL} and \textsc{HUD} ($p = 0.009$). No other significant differences were indicated. Hence, further descriptive statistics are omitted for brevity. \rev{In other words, \textsc{DUAL} led to a significantly higher mental demand compared to both \textsc{Baseline} and \textsc{HUD}.}


\textbf{Situation Awareness:} 

\begin{table}
\caption{Average SART results for Experiment 3 with standard deviation in parenthesis. SA: Spatial Awareness. For SA, supply and understanding, higher scores are better, for demand, lower scores are better. Grey rows and bold numbers indicate scales with significant differences.}

\begin{center}
    \small
	\begin{tabular}{ |c|c|c|c| }
            \hline 
		Scale & \textsc{Baseline} & \textsc{HUD} & \textsc{DUAL} \\
            \hline 
		SA & 14.6  & 12.2  & 14.0\\
		& (4.0) &  (4.4) & (4.6) \\
            \hline 
            \rowcolor{lightgray}
		Demand & \textbf{9.6} & \textbf{12.8} & \textbf{10.66}\\
            \rowcolor{lightgray}
		& \textbf{(3.3)} &  \textbf{(3.3)} & \textbf{(3.1)} \\
            \hline 
		Supply & 16.5 & 18.5 & 17.5  \\
		& (4.0) & (3.7) & (4.0) \\
            \hline 
		Understanding & 7.7 & 6.5 & 7.0 \\
		& (2.1) & (2.7) & (1.6) \\
            \hline 
	\end{tabular}
\end{center}
\label{tab:sarte3}
\end{table}


\rev{SART scores are depicted in Table \ref{tab:sarte3}. Friedman's tests revealed a statistically significant difference for demand subscale ($\chi^2(2) = 10.05$, $p = 0.007$), but not for the overall SART score or the understanding and supply subscales. For demand, post-hoc analysis with Wilcoxon signed-rank tests and Holm-Bonferroni correction revealed that there were significant differences between \textsc{HUD} and \textsc{baseline} ($p = 0.002$), as well as \textsc{HUD} and \textsc{DUAL} ($p = 0.008$). In other words, \textsc{HUD} led to a significantly higher demand compared to both \textsc{baseline} and \textsc{DUAL}.}



\begin{table}
\caption{Average SSQ results for Experiment 3 with standard deviation in parenthesis. Grey rows and bold numbers indicate scales with significant differences.}
\begin{center}
    \small
 	\begin{tabular}{ |c|c|c|c| }
              \hline 
		Scale & \textsc{Baseline} & \textsc{HUD} & \textsc{Dual} \\
              \hline 
              \rowcolor{lightgray}
		Total & \textbf{36.5}  & 45.3  & \textbf{47.8} \\
  \rowcolor{lightgray}
		& \textbf{(36.0)} & (38.9) & \textbf{(41.1)} \\
              \hline 
  \rowcolor{lightgray}
		Nausea & \textbf{46.1} & 58.4 & \textbf{58.8}\\
  \rowcolor{lightgray}
		& \textbf{(38.8)} & (44.8) & \textbf{(43.4)} \\
              \hline 
		Oculo-Motor & 17.1 & 19.3 & 21.2  \\
		& (21.4) & (25.2) & (22.9) \\
              \hline 
  \rowcolor{lightgray}
		Disorientation & \textbf{37.1} & 48.1 & \textbf{53.4} \\
  \rowcolor{lightgray}
		& \textbf{(47.5)} & (47.3) & \textbf{(58.4)} \\
              \hline 
	\end{tabular}
 \vspace{-0.5cm}
\end{center}
\label{tab:ssqe3}
\end{table}

\textbf{Simulator Sickness: } \rev{SSQ scores are depicted in Table} \ref{tab:ssqe3}. 
\rev{Friedman's tests revealed statistically significant differences for nausea ($\chi^2(2) = 8.03$, $p = 0.018$), disorientation ($\chi^2(2) = 7.05$, $p = 0.029$) and total severity ($\chi^2(2) = 10.05$, $p = 0.007$).} 


\rev{Post-hoc analysis with Wilcoxon signed-rank tests and Holm-Bonferroni correction revealed that there were significant differences between \textsc{Dual} and \textsc{baseline} for nausea ($p = 0.007$), for disorientation ($p = 0.003$) and for total severity ($p = 0.004$). No other significant differences were indicated. In other words, \textsc{Dual} led to significantly higher SSQ scores (except oculo-motor) compared to \textsc{Baseline}. The symptoms can be regarded as slight.}



\textbf{Flow: } 
The mean overall score for flow as measured by FSS  was 4.75 (sd = 0.63) for \textsc{Baseline}, 4.58 (sd = 0.67) for \textsc{HUD} and 4.21 (sd = 0.79) for \textsc{Dual}. For anxiety, it was 2.49 (sd = 1.116) for \textsc{Baseline}, 3.03 (sd = 1.55) for \textsc{HUD} and 2.71 (sd = 1.27) for \textsc{Dual}. For challenge, it was 3.96 (sd = 0.55) for \textsc{Baseline}, 4.17 (sd = 0.48) for \textsc{HUD} and 4.46 (sd = 0.83) for \textsc{Dual}. 

Friedman's tests revealed statistically significant differences for the flow ($\chi^2(2) = 7.298$, $p = 0.026$) and challenge ($\chi^2(2) = 6.045$, $p = 0.049$) subscales, but not for anxiety. For flow, post-hoc analysis with Wilcoxon signed-rank tests and Holm-Bonferroni correction revealed that there was a significant difference between \textsc{HUD} and \textsc{baseline} ($p = 0.013$). No other significant differences were indicated. \rev{In other words, \textsc{HUD} led to a significantly lower flow score (12\%) compared to \textsc{Baseline}.}


\textbf{Preferences and Open Comments: }

Participants were asked to rank the conditions from least preferred to most preferred. Nineteen out of 24 participants preferred \textsc{Baseline}, two \textsc{HUD} and three \textsc{Dual}.

\rev{Friedman's tests revealed statistically significant differences on ranks} ($\chi^2(2) = 22.6$, $p < 0.001$). \rev{Post-hoc analysis revealed that there were significant differences between \textsc{baseline} and \textsc{HUD} }($p < 0.001$) \rev{and \textsc{baseline} and \textsc{DUAL} }($p < 0.001$), \rev{but not between \textsc{HUD} and \textsc{DUAL}}.

Five participants explicitly highlighted that they could concentrate best in the smartphone-only condition, with one mentioning ``The fewer devices I had, the better I could concentrate'' and another ``I was least overwhelmed by stimuli''.
One participant explicitly mentioned, that she ``needed to switch attention less often, as keyboard and text output were spatially very close'' and another one ``using only the smartphone, I can still see my feet.''
However, one participant also mentioned ``I tend to forget to concentrate on the environment [using solely the smartphone]'' and another related to nausea when using the smartphone while walking: ``I got sick when using the smartphone, using the HMD along with the smartphone reduced this sickness.''

Participants also mentioned the benefits of using the HMD, e.g. ``reading on OST HMD was more comfortable'' and ``reading on the HMD made me more aware of the environment''. One participant explicitly stated that the \textsc{HUD} condition was ``best for writing and [simultaneous] walking''. On the other hand, comments addressing problematic effects included ``switching between depth layers was annoying'', and three participants mentioned readability issues on the HMD. Also, participants mentioned the need for context switching resulting in higher workload with comments such as ``I needed to reorient myself constantly'', ``repeated switching between smartphone and HMD lead to concentration errors'', or ``the attention I gain for the environment is mitigated by the attention I loose for text input.'' One participant would have liked haptic feedback on the smartphone keyboard to help with writing.

Additionally, for condition \textsc{Dual} participants mentioned, that they used the HMD less often with comments such as ``I barely looked at the HMD'' or ``This condition leads to irritation where to actually look''. However, one participant also stated ``It is relieving to be able to read the text both on the smartphone as well as in front of you.''









\subsection{Experiment 4: Obstacle Collisions}


Experiment 3 used a physical obstacle course and revealed that participants looked significantly less at the smartphone in the \textsc{HUD} condition \rev{at the expense of text entry performance and attentional demand}. Subjective feedback from participants indicated a potentially higher awareness of the environment (which was not reflected in SART scores). As eye gaze alone is not a suitable indicator of true attention to the environment (participants could have solely focused their attention on the virtual text in the HMD), we set out to further investigate this potential factor in another experiment. As participants tended to not make use of the HMD in the  \textsc{Dual} condition in Experiment 3, we removed this condition for Experiment 4. \rev{Instead of a physical course, a virtual obstacle course was set up. This was done for two reasons. First, the physical obstacle course led to a frequent change in walking speed (starting to walk and stopping again after a few meters). In addition, we wanted to investigate behavior on an extended straight boardwalk. Second, we wanted to mitigate the risks of injuries when running into physical obstacles.
}





\subsubsection{Method}
The experiment was a within-subjects design with one independent variable \textsc{interface}. The independent variable \textsc{interface} had two levels: \textsc{Baseline} and \textsc{HUD}, which were the same conditions used in Experiment 2. The order of the starting conditions was balanced across participants. The task was to write as quickly and as accurately as possible while walking on a physical treadmill and avoiding virtual obstacles. A screen in front of the participant showed a virtual path moving at the same speed (1.5 km/h) as the physical treadmill. On that path, virtual obstacles appeared in the middle, to the left, and to the right of the participant. The participant was asked to indicate if they noticed the virtual obstacles in the middle of the path  (as they could bump into them) with a button attached to the back of the smartphone. \mr{The button press was registered to the nearest obstacle. On average, the participants had between 5 and 19 seconds to react to the closest visible obstacle.} 

\subsubsection{Participants}
We recruited 17 volunteers. None had participated in the previous experiments. Three volunteers had to abort the study due to incompatibility with the HMD. Those participants experienced visual problems (diplopia, focussing issues) with the HMD during the experiment. For the remaining 14 volunteers (mean age 26.71 years, sd = 3.36, mean height 174.36 cm, sd = 8.41, 8 male, 6 female), six indicated to have visual restrictions. Three used contact lenses to correct these. No participant used additional glasses during the experiment. All participants were familiar with typing on a smartphone. Eight participants never used an HMD before, one once, four rarely, and one often.

\subsubsection{Apparatus and Materials}
\begin{figure}[tb]
	\centering
	 \includegraphics[width=\columnwidth]{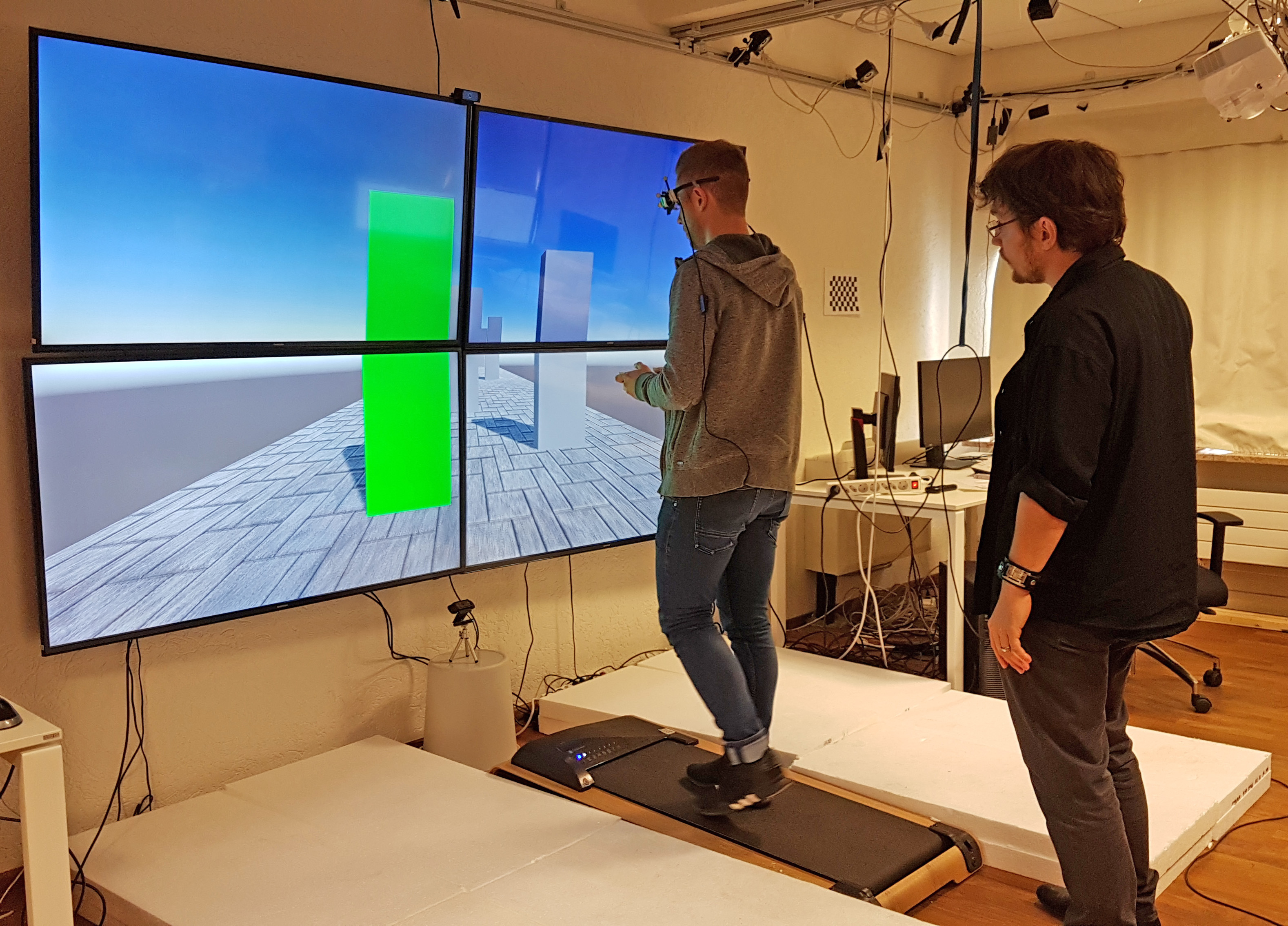}
	\caption{Setup for Experiment 4, obstacle collisions: A participant (middle person) being secured by an experimenter (right person) and use of styrofoam on the side to prevent potentially slipping off the treadmill. On the screen in front of the participant, an obstacle has been triggered as detected (highlighted in green) by the participant pushing a button on the back of the smartphone. }	\label{fig:apparatustreadmill}
\end{figure}

The study setup is shown in Figure \ref{fig:apparatustreadmill}. For this study, an Epson Moverio BT-300 was used along with a Moto Z2 Play smartphone for text entry. The HMD was equipped with a Pupil Labs mobile eye-tracker (attached to a stationary PC for recording) and with retro-reflective markers. An Optitrack V120:Trio system was used for tracking of the head tilt. Additionally, cameras were mounted to the side and in front of the participants. The treadmill was a DeskFit200 with conveyor belt dimensions of 40x90.5 cm.  The treadmill was secured with styrofoam on the sides and one experimenter was positioned behind the participants to address participants potentially slipping off the treadmill. Four Samsung UE55MU6179 4K monitors were used together as a single large display with the dimensions 248 $\times$ 143 cm and were placed 100 cm in front of the participant. This resulted in an approximate horizontal field of view of the virtual path of 102\textdegree~and vertical 71\textdegree.
The obstacles were placed on an endless virtual path which had a width of 5 meters. Obstacles with 60 cm width appeared in three distinct lanes (left - 1.5 m from the center, middle, right - 1.5 m from the center). Obstacle frequency and placement were randomized for each lane, with all lanes using the same parameters, which were as follows: a new obstacle was placed after an average of 5 m, with the distances offset using a uniformly distributed random value in a range of 2 m to 8 m. Center obstacles were also randomly offset to the left or right by 40 cm. Outer obstacles were offset randomly in a range of 20 cm to the center, and also uniformly distributed.

\subsubsection{Procedure}
Participants were asked to fill out a demographic questionnaire and then tested to ensure they were able to read text on the HMD using a test text. Participants' heights were used to adjust the virtual camera height for the obstacle course. 
Participants were then asked to get used to the treadmill in a 2-minute test run without wearing the HMD or smartphone in order to familiarize themselves with the equipment. We calibrated the eye tracker and conducted a 5-minute testing phase before each condition. When a condition had been completed participants filled out the same questionnaires as in Experiment 2 (TLX, SSQ, SART, FSS) and were given an additional 5 minutes for resting. After the final condition, participants were asked to fill out a self-assessment questionnaire, which was followed by a semi-structured interview. 
The order of the starting conditions was balanced across participants. In either condition, participants were shown a series of stimulus sentences. The experiment was carried out in a 75-minute session. Starting with a 20-minute introduction, briefing phase, and calibration followed by a 45-minute testing phase (5 minutes calibration per condition + 15 minutes per condition + 10 minutes questionnaires + 5-minute break between conditions) and 10 minutes for the final questionnaire, interview, and debriefing.

\subsubsection{Results}

\rev{In addition to significance tests, we also conducted two one-sided (paired t-)test (TOST) for equivalence testing with equivalence bounds of $d_z=\pm0.56$ (based on a critical t-value of 2.16
at an alpha value of 0.05 and a sample size of 14
, c.f. \cite{lakens2013calculating, lakens2018equivalence}.}


\textbf{Obstacle collisions: }
We measured the collision ratio (between 0 and 100\%) as the number of undetected collisions relative to the overall number of possible collisions. For \textsc{Baseline}, the mean collision ratio was 19.84\% (sd = 11.43) and for \textsc{HUD}, it was 19.35\% (sd = 14.67). We also measured a false positive ratio, i.e. the number of falsely identified obstacles (e.g., obstacles left or right of the user). For \textsc{Baseline}, the false positive ratio was 23.64\% (sd = 9.55) and for \textsc{HUD}, it was 24.10\% (sd = 10.53). Paired two-tailed t-tests did not reveal a significant difference ($t(13) = 0.257, p = 0.8$). \rev{TOST indicated equivalence between conditions for both collision ratio (results for the larger of the two p-values: ($t(13) = 1.84, p = 0.045$) and false positive ratio ($t(13) = -1.83, p = 0.045$). In other words, both conditions were equivalent in terms of obstacle collisions.}

\textbf{Eye Gaze and Head Tilt: }
We investigated the duration participants spent looking at the smartphone. Data from three participants had to be excluded due to non-working eye-tracking. The mean relative duration for \textsc{Baseline} was 38.70\% (sd = 21.05) and for \textsc{HUD} 24.67\% (sd = 12.00). Paired two-tailed t-tests did reveal a significant difference ($t(10) = 3.172, p = 0.10$).

We also measured the head tilt of participants, with 0\textdegree~looking straight ahead and -90\textdegree~looking straight down. For \textsc{Baseline}, the mean head tilt was -39.12\textdegree~(sd = 46.64) and for \textsc{HUD}, it was -28.58\textdegree~(sd = 42.13). Paired two-tailed t-tests did not reveal a significant difference. \rev{In other words, participants looked significantly less on the smartphone in the \textsc{HUD} condition compared to \textsc{Baseline}.}

\textbf{Text Entry Rate and Error Rate: }
For \textsc{Baseline}, the mean entry rate was 13.84 wpm (sd = 4.09) and for \textsc{HUD} 12.84 wpm (sd = 5.46). The mean character error rate was 0.018 (sd = 0.023) for \textsc{BASELINE} and for \textsc{HUD}, it was 0.045 (sd = 0.070). Paired two-tailed t-tests did not reveal any significant differences. \rev{Also, no equivalence between conditions could be detected. In other words, no significant differences in text entry or error rate could be detected, but the conditions cannot be seen as equivalent.}

\textbf{Workload: }
The mean overall score for workload as measured by Nasa TLX was 58.45 (sd = 9.37) for \textsc{baseline} and 61.67 (sd = 11.05) for \textsc{HUD}. Wilcoxon signed-rank tests did not reveal statistically significant differences for the overall demand score  ($p > 0.05$) or any of the subscales, which are omitted for brevity. \rev{Also, no equivalence between conditions could be detected. In other words, no significant differences in workload could be detected but conditions cannot be considered equivalent.}


\textbf{Situation Awareness: } Descriptive statistics for SART scores are depicted in Table \ref{tab:sarte4}. Wilcoxon signed-rank tests did not reveal statistically significant differences for the overall SART score  ($p > 0.05$) or any of the subscales, which are omitted for brevity. \rev{Also, no equivalence between conditions could be detected. In other words, no significant differences in situation awareness could be detected, but conditions cannot be considered equivalent.}

\begin{table}
\caption{Average SART results for Experiment 4 with standard deviation in parenthesis. SA: Spatial Awareness. For SA, supply and understanding, higher scores are better, for demand, lower scores are better. }
\begin{center}
	\small
	\begin{tabular}{ |c|c|c| }
	    \hline 
		Scale & \textsc{Baseline} & \textsc{HUD} \\
            \hline
		SA & 13.9  & 11.9\\
		& (5.1) &  (5.6)  \\
            \hline
		Demand & 12.1  & 13.2\\
		& (3.4) &  (3.9)  \\
              \hline
		Supply & 19.4 & 19.5  \\
		& (3.3) & (3.5)  \\
              \hline
		Understanding & 6.6 & 5.6 \\
		& (2.3) & (2.1) \\
              \hline
	\end{tabular}
\end{center}
\label{tab:sarte4}
\vspace{-0.5cm}
\end{table}


\textbf{Simulator Sickness: }
The scores for the SSQ scales are depicted in Table \ref{tab:ssqe4}. \rev{Wilcoxon signed-rank tests revealed significant differences between \textsc{Baseline} and \textsc{HUD} for disorientation ($Z = 2.36, p = 0.018$) and total severity ($Z = 2.11, p = 0.035$) but not for nausea or oculo-motor. In other words, \textsc{HUD} led to significantly higher disorientation and total severity scores than the baseline condition, but those scores can still be considered to indicate mild symptoms.}

\begin{table}
\caption{Average SSQ results for Experiment 4 with standard deviation in parenthesis. Grey rows and bold numbers indicate scales with significant differences.}

\begin{center}
    \small
	\begin{tabular}{ |c|c|c| }
              \hline
		Scale & \textsc{Baseline} & \textsc{HUD} \\
              \hline
              \rowcolor{lightgray}
		Total & \textbf{17.9}  & \textbf{26.2}   \\
            \rowcolor{lightgray}
		& \textbf{(18.6)} &  \textbf{(19.9)}  \\
              \hline
              
		Nausea & 22.5 & 33.4 \\
  
		& (21.0) & (23.9)  \\
            
              \hline
		Oculo-Motor & 11.4 & 12.5   \\
		& (10.6) & (10.6)  \\
              \hline
              \rowcolor{lightgray}
		Disorientation & \textbf{12.9} & \textbf{25.9}  \\
            \rowcolor{lightgray}
		& \textbf{(22.2)} & \textbf{(28.3)} \\
              \hline
              
	\end{tabular}
 \vspace{-0.5cm}
\end{center}

\label{tab:ssqe4}
\end{table}



\textbf{Flow: }
The mean overall score for flow as measured by FSS  was 3.58 (sd = 2.02) for \textsc{Baseline} and 3.35 (sd = 2.09) for \textsc{HUD}. For anxiety, it was 2.62 (sd = 2.34) for \textsc{baseline} and 2.69 (sd = 2.34) for \textsc{HUD}. For challenge, it was 3.36 (sd = 1.865) for \textsc{Baseline} and 3.43 (sd = 1.95) for \textsc{HUD}. Wilcoxon signed-rank tests did not reveal statistically significant differences \rev{ and TOST did not indicate equivalence. In other words, no significant differences between the conditions could be detected for flow.} 


\textbf{Preferences and Open Comments: }

Participants were asked to rank the conditions from least preferred to most preferred. Nine out of 14 participants preferred \textsc{baseline}, five the \textsc{hud} condition. \rev{A binomial test did not indicate a significant difference ($p = 0.424$)}. 
Two participants mentioned that using only the smartphone led to more collisions. “I ran into more obstacles because I was fixated on the smartphone.” For the benefits of using only the smartphone, three participants said, that they are used to writing and reading on it. “I am using it daily. I can write without looking at it.” Two participants mentioned, that they had difficulty switching between the different depth layers. “It’s hard for me to adjust my focus on three different things. It’s inconvenient.” Four participants mentioned that they had a hard time getting a clear view of the text in the HMD: “The text was blurry until I concentrated hard on reading it,” “My eyes needed time to adjust to HMD.” This was reconfirmed by two participants mentioning that their eyes were strained after the HMD condition. One participant stated that the HMD display obstructed the view of the obstacles: “I couldn’t see the obstacles because the projection was in
the way.” As for the benefits of the HMD, five participants mentioned that the HMD allowed them to see more of their surroundings, with one participant saying “It feels that I can see more of my surroundings and the obstacles with the HMD.” Two participants also mentioned that they feel safer using it: “I was maybe slower with the HMD, but I feel safer.” Two participants imagined that further practice with the HMD could lead to better results: “Given more time and getting used to the HMD, I could have avoided more obstacles.”


\subsection{\rev{Experiment 5: Effects of Wearing an HMD}}


\rev{As prior research has indicated that wearing an HMD could lead to perceptual effects, for example, on distance perception} \cite{messing2005distance} \rev{due to a restricted field of vision} \cite{darius2015investigation}\rev{, we investigated if there is an effect of wearing an  HMD (that is turned off) on obstacle avoidance. To this end, we reran Experiment 4, but with the following changes:}

\rev{The independent variable \textsc{interface} had two levels: \textsc{Baseline} and \textsc{HMD-OFF}. \textsc{HMD-OFF} used the same input and output channels (the smartphone) as \textsc{Baseline}, but, in addition, required users to wear the same HMD as in condition \textsc{HUD}, which was turned off. In the procedure, we omitted eye tracking as the input and output happened on the same device.} 

\rev{Fourteen volunteers participated in the study (mean age 27.5 years sd = 8.5, 10 male, 4 female). All participants were familiar with typing on a smartphone. Six participants had never worn an HMD before, 4 once, 2 rarely, 1 occasionally, and 1 frequently.}

\subsubsection{\rev{Results}}
\rev{In addition, to significance tests, we also conducted a two one-sided paired t-test for equivalence testing 
with equivalence bounds of $d_z=\pm0.58$ (based on a critical t-value of 2.16, an alpha value of 0.05 and a sample size of 14. Data from two participants for the challenge subscale of the flow questionnaire were lost due to logging errors.}

\textbf{\rev{Obstacle collisions: }}
 \rev{For \textsc{Baseline}, the mean collision ratio was 15.6\% (sd = 6.5) and for \textsc{HMD-OFF} it was 20.0\% (sd = 11.0). For \textsc{Baseline}, the false positive ratio was 19.9\% (sd = 8.9) and for \textsc{HMD-OFF}, it was 21.7\% (sd = 12.2). Paired two-tailed t-tests did not reveal a significant difference and TOST did not indicate equivalence.}

\textbf{\rev{Text Entry Rate and Error Rate: }}
\rev{
For \textsc{Baseline}, the mean entry rate was 14.8 wpm (sd = 4.4) and for \textsc{HMD-OFF} 14.7 wpm (sd = 3.8). The mean character error rate was 0.016 (sd = 0.020) for \textsc{BASELINE} and for \textsc{HMD-OFF}, it was 0.014 (sd = 0.013). %
TOST did indicate equivalence for text entry rate (results for the larger of the two p-values: ($t(13) = 1.771, p = 0.005$)) but not character error rate. 
To summarize, the text entry but not the error rate was equivalent between both conditions.}


\textbf{\rev{Workload: }}
\rev{
The mean overall score for workload as measured by Nasa TLX was 62.02 (sd = 13.93) for \textsc{Baseline} and 61.37 (sd = 12.53) for \textsc{HMD-OFF}. Wilcoxon signed-rank tests did not reveal statistically significant differences for the overall demand score  ($p > 0.05$) or any of the subscales, which are omitted for brevity. TOST indicated equivalence for physical demand (results for the larger of the two p-values: $t(13) = 1.86, p = 0.043$), frustration ($t(13) = -1.99, p = 0.034$) and overall demand ($t(13) = 1.86, p = 0.043$), but not for mental demand, temporal demand, effort or performance. In other words, the conditions can be seen as equivalent in terms of overall demand (but not for each subscale).
}

\textbf{\rev{Situation Awareness: }} \rev{SART results are depicted in Table} \ref{tab:sarte5} \rev{Wilcoxon signed-rank tests indicated a significant difference for the supply subscale (i.e. how users can supply their attentional resources using items on arousal, spare mental capacity, concentration and division of attention) between \textsc{Baseline} and \textsc{HMD-OFF} (Z = 2.17, p = 0.03, Cohen's d = 0.9), but not for demand, understanding or overall. Also, no equivalence between conditions for any of those measures was indicated. In other words, wearing an HMD, even though it was turned off led to a significantly lower supply rating compared to \textsc{Baseline}.}




\begin{table}
\caption{Average SART results for Experiment 5 with standard deviation in parenthesis. SA: Spatial Awareness. For SA, supply, and understanding, higher scores are better, for demand, lower scores are better. Grey rows and bold numbers indicate significant differences.}
\begin{center}
        \small
	\begin{tabular}{ |c|c|c| }
            \hline
		Scale & \textsc{Baseline} & \textsc{HMD-OFF} \\
            \hline
		SA & 14.3  & 12.4\\
		& (5.9) &  (5.9)  \\
            \hline
		Demand & 11.9  & 12.4\\
		& (3.7) &  (4.0)  \\
            \hline
            \rowcolor{lightgray}
		Supply & \textbf{19.1} & \textbf{17.4}  \\
            \rowcolor{lightgray}
		& \textbf{(4.1)} & \textbf{(4.9)}  \\
            \hline
		Understanding & 7.1 & 7.5 \\
		& (2.6) & (2.3) \\
            \hline
	\end{tabular}
\end{center}
\label{tab:sarte5}
\end{table}


\textbf{\rev{Simulator Sickness: }}
\rev{The scores for the SSQ scales are depicted in Table} \ref{tab:ssqe5}\rev{. TOST indicated equivalence for nausea (results for the larger of the two p-values: $t(13) = 2.10, p = 0.028$), disorientation $t(13) = 2.10, p = 0.028$), and total severity ($t(13) = -197, p = 0.035$), but not for oculo-motor ($t(13) = -165, p = 0.061$). In other words, both conditions can be considered equivalent in terms of induced simulator sickness for all but the oculo-motor component.}



\begin{table}
\caption{Average SSQ results for Experiment 5 with standard deviation in parenthesis.}
\begin{center}
    \small
	\begin{tabular}{ |c|c|c| }
            \hline
		Scale & \textsc{Baseline} & \textsc{HMD-OFF} \\
            \hline
		Total & 35.0  & 35.8   \\
		& (41.3) &  (35.3)  \\
            \hline
		Nausea & 42.2 & 42.2 \\
		& (49.3) & (40.9)  \\
            \hline
		Oculo-Motor & 20.0 & 21.7   \\
		& (19.6) & (19.9)  \\
            \hline
		Disorientation & 31.8 & 31.8  \\
		& (50.2) & (46.1) \\
            \hline
	\end{tabular}
\end{center}
\label{tab:ssqe5}
\vspace{-0.5cm}
\end{table}



\textbf{\rev{Flow: }}
\rev{The mean overall score for flow as measured by FSS  was 4.57 (sd = 1.11) for \textsc{Baseline} and 4.61 (sd = 1.02) for \textsc{HMD-OFF}. For anxiety, it was 3.74 (sd = 1.50) for \textsc{baseline} and 2.95 (sd = 1.51) for \textsc{HMD-OFF}. For challenge, it was 4.67 (sd = 0.99) for \textsc{Baseline} and 4.92 (sd = 0.90) for \textsc{HMD-OFF}. Wilcoxon signed-rank tests did not reveal statistically significant differences. TOST indicated equivalence for flow (results for the larger of the two p-values: 
$t(13) = -3.44, p = 0.002$), but not for anxiety and challenge. In other words, no significant differences between conditions could be detected for flow.
}

\textbf{\rev{Open Comments: }}

\rev{In a semi-structured interview, participants were asked if they felt that wearing the HMD (even though it was turned off) was impacting the task. 
Seven participants did not notice a difference between \textsc{Baseline} and \textsc{HMD-OFF}. Five participants mentioned that \textsc{HMD-OFF} was negatively impacting the task. All of those five participants usually do not wear corrective glasses. Two participants mentioned that the HMD was limiting the available field-of-view. Another one mentioned that wearing the HMD felt unusual. Yet, another two participants explicitly mentioned, that they felt that they were recognizing the targets less often when wearing the HMD, with one mentioning looking through the HMD "felt like looking through a veil".
Two participants, preferred wearing the HMD, with one mentioning a learning effect (the participant conducted the \textsc{HMD-OFF} condition after \textsc{Baseline}) and another one mentioning fatiguing effects (\textsc{HMD-OFF} was conducted before \textsc{Baseline}) as reasons for this preference.
}

\section{Discussion}
In this paper, we have studied the possible merits and challenges of a joint OST HMD-smartphone system for mobile text entry. 

\rev{In Experiment 1, we quantified the performance gains a full-screen smartphone keyboard could have over a  standard-size smartphone keyboard. We verified that the full-screen keyboard lead to a significantly higher text entry rate (around 15\%) with no significant difference in error rate. In addition, the larger keyboard also led to a lower workload, which was also supported by participants' remarks.}

\rev{In Experiment 2, we found that wearing an HMD for text output significantly reduced text entry performance compared to a standard smartphone baseline. In addition, we found that a spatially registered AR text output view performed worse than a HUD text output view. This result is supported by the literature on context and focus distance switching (e.g., \cite{eiberger2019effects, arefin2022effect}), which indicate substantial costs when visual information is processed across an OST HMD and another display. Specifically, Eiberger et al. \cite{eiberger2019effects} studied a setup very similar to ours, with a comparable HMD model and a secondary display at a typical reading distance. While we could not isolate the effects of the joint visual information processing in experiment 2, the overall results, together with the results from experiment 1 (higher text entry performance with full-screen keyboard), and results reported in prior work, it seems clear that there are major costs introduced due to the need for switching between the HMD and the smartphone. \mr{In addition, while we ensured that the \textsc{AR} condition was spatially registered for each participant, we did not use a user-specific calibration. In future work, one should further investigate that factor by either utilizing user-specific calibrations or replicating the experiment with a high-resolution video see-through display.} \mr{Also, while previous work \cite{grubert2015multifi} considered joint OST HMD-smartphone systems having potential benefits for supporting privacy, the subjective feedback in Experiment 2 did not confirm this.}} 

\rev{
Due to observations that participants looked down at the smartphone considerably less in Experiment 2, we hypothesized that there could be potential advantages in mobile scenarios in which users need to divide attention between the interactive system and the surrounding environment.
Hence, we ran another set of experiments focusing on the need for splitting the attention between text entry and the physical environment. To this end, Experiment 3 used a physical obstacle course and Experiments 4 and 5 used a treadmill and virtual obstacles to compare typing using the joint OST HMD-smartphone system with typing on a smartphone only. 
}

\rev{
Experiment 3 (physical obstacle course) showed that typing with a joint OST HMD-smartphone system led to significantly slower and more error-prone text entry compared to smartphone-only typing. This experiment also indicated that users tended to use the smartphone if the text was simply mirrored on the HMD. Subjective feedback indicated that participants experienced significantly more oculo-motor symptoms when wearing the HMD and most participants (19 out of 24) preferred smartphone-only typing. However, this was not reflected in the overall workload or flow ratings. Also, we found that users tended to look on the smartphone about 70\% of the time when using the smartphone for text input and output compared to approximately 40\% of the time when using the HMD for text output. This was accompanied by user comments on increased awareness of the environment (even though this was not reflected in SART scores).
}

\rev{
However, the design of Experiment 3 meant that we were unable to detect whether participants' awareness had actually improved as additional attention might have been simply directed towards the HMD. In addition, the physical obstacle course did not allow us to study extended continuous walking as participants needed to slow down and change direction after a few meters.}
\rev{
Therefore Experiment 4 used virtual obstacles on a treadmill, which allowed us to investigate whether participants were able to avoid obstacles less with an HMD. 
}
\rev{
Experiment 4 revealed that the obstacle collision rate was equivalent (around 20~\%) for both conditions and that the \textsc{HUD} condition resulted in significantly higher (but still mild) simulator sickness in terms of disorientation and total severity. 
}

\rev{
Finally, based on prior indications \cite{darius2015investigation}, Experiment 5 investigated whether simply wearing an OST HMD that is turned off would have an impact on user behavior in the same task as in Experiment 4. The results indicated equivalence for text entry speed, overall demand, and simulator sickness (except the oculo-motor subscale). At the same time, wearing an HMD led to a significantly lower attentional supply. 
}

In summary, the \rev{five} experiments have contributed to a nuanced understanding of the performance of a joint OST HMD-smartphone system. In a static setting, the joint OST HMD-smartphone system has significantly reduced performance compared to a smartphone-only baseline. In a dynamic setting, where users must maintain awareness of their surroundings, the joint OST HMD-smartphone system reduces the number of times users look at the smartphone. However, users are still not further aware of their surroundings as evidenced by the obstacle collision rate being similar for both systems. \rev{Hence, further research is needed to investigate if it is possible, and if so how, to translate the theoretical performance benefits of a joint OST HMD-smartphone system into mobile settings.}

\rev{
For example, the setup used a current-generation OST HMD with a single focal plane. Prior work has demonstrated the costs of focal plane switching (e.g. \cite{eiberger2019effects, arefin2020impact, arefin2022effect}), which might also have impacted text entry performance in our study. Possibly, varifocal or multiple foci OST HMDs \cite{dunn2017wide, akcsit2017near, wilson2018high, itoh2021towards} would reduce this negative impact by aligning both the focal plane of the smartphone and the HMD or of the HMD and the physical environment (if users are looking at the physical surroundings). Still, even if this could be realized, there would still remain costs of context switching \cite{huckauf2010perceptual, arefin2020impact, arefin2022effect}. Hence, it remains to be seen if future OST HMD systems could substantially improve the situation awareness or text entry performance. Until then, we would caution against proposing OST HMDs as a likely solution for reducing distractions for mobile text entry or other smartphone-related activities demanding visual attention.}




Our experiments could be modified by changing the parameters, such as changing the frequency of obstacles. It is also possible to change the nature of the obstacles, such as having obstacles on the ground. While it is possible to conceive many alternative experimental designs, on balance, we believe our results are robust to minor adjustments in the design and procedure. Further experiments could consider alternative ways of presenting text on an HMD \cite{rzayev2018reading}.

While it was not viable to increase exposure time in our experimental setups, we did get indications from some participants that they could possibly increase their performance with practice. In general, we suggest an interesting avenue for future work would be to study OST HMD-smartphone systems, and other complex designs, longitudinal in real interaction contexts, possibly using the Experience Sampling Methodology  \cite{consolvo2003using}.

\rev{Finally, the studies were relatively short interventions. Hence, we could not investigate long-term learning effects. Quantifying potentially different learning rates and performance bounds is one avenue of future work.}

\section{Conclusions}

The central contribution of this paper is a series of investigations that provide a nuanced understanding of the opportunities and challenges inherent in a joint OST HMD-smartphone system. To date, OST HMDs lack efficient text entry methods. Since smartphones are a major text entry medium in mobile contexts but their attentional demands can contribute to accidents while typing on the go, we explored the possible performance benefits of a joint OST HMD-smartphone system. In a series of five experiments with a total of 86 participants we found that, as of today, the challenges inherent in a joint OST HMD-smartphone system outweigh the potential benefits compared to a smartphone-only baseline. The experiments confirmed previous findings about the performance issues in multi-depth information environments that come along with today's single-focus HMD design. We hope that joint HMD-smartphone ecologies get a boost from upcoming multi-focus and variofocal HMD designs.

\bibliographystyle{IEEEtran}
%

\bibliography{arte}




%
\begin{IEEEbiography}[{\includegraphics[width=1in,height=1.25in,clip,keepaspectratio]{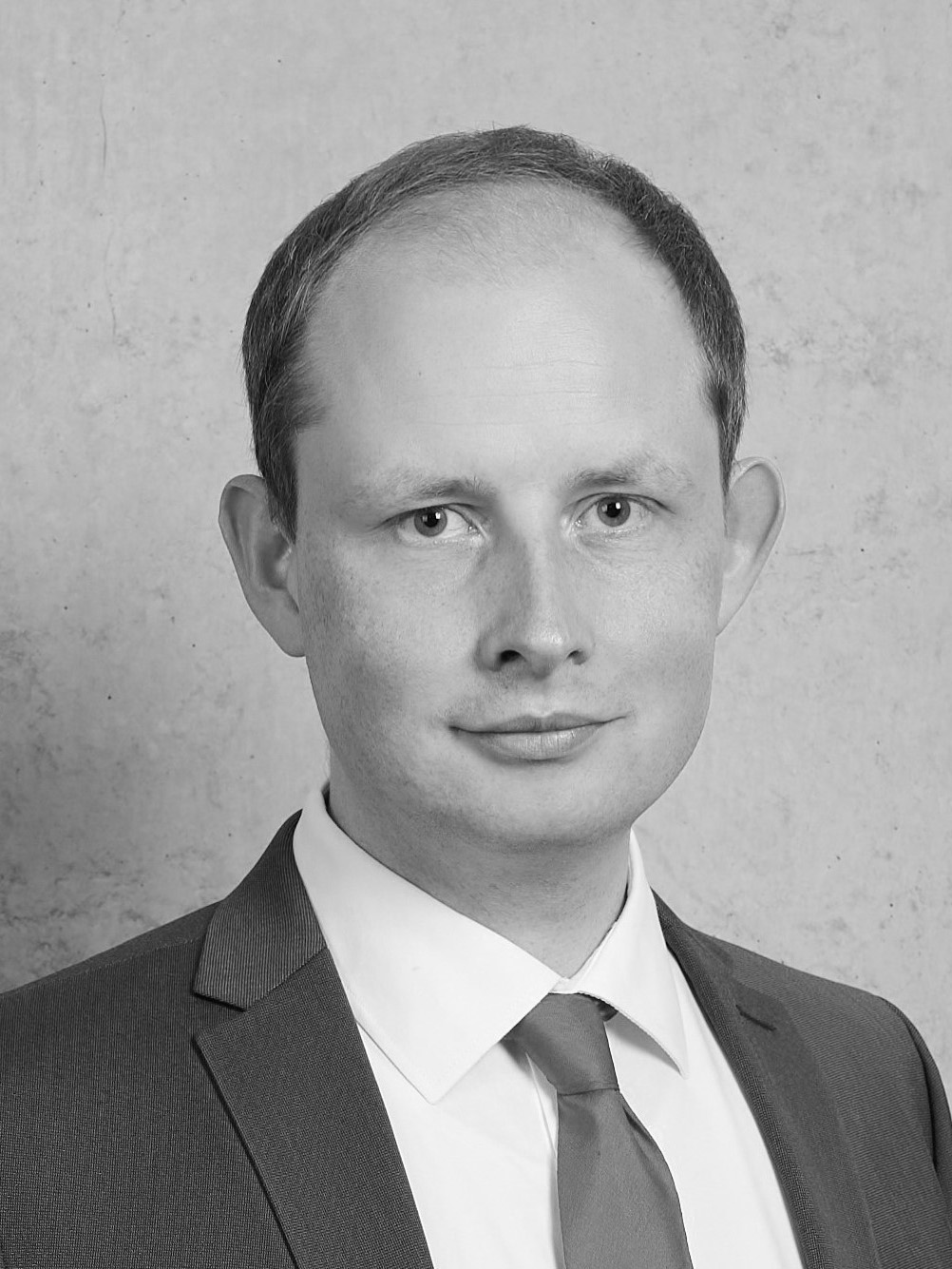}}]{Jens Grubert}
 is Professor of Human-Computer Interaction in the Internet of Things at Coburg University of Applied Sciences and Arts. He has more than 15 years of academic and industrial experience in fields such as human-computer interaction, mixed reality, computer graphics, and computer vision. His research interests include supporting future knowledge work with mixed reality, multimodal and mutli-device interaction in augmented and virtual reality. He is an avid supporter of interdisciplinary work.
\end{IEEEbiography}


\begin{IEEEbiographynophoto}{Lukas Witzani}
graduated from the University of Passau where he was conducting research on text entry in virtual reality.
\end{IEEEbiographynophoto}

\begin{IEEEbiography}[{\includegraphics[width=1in,height=1.25in,clip,keepaspectratio]{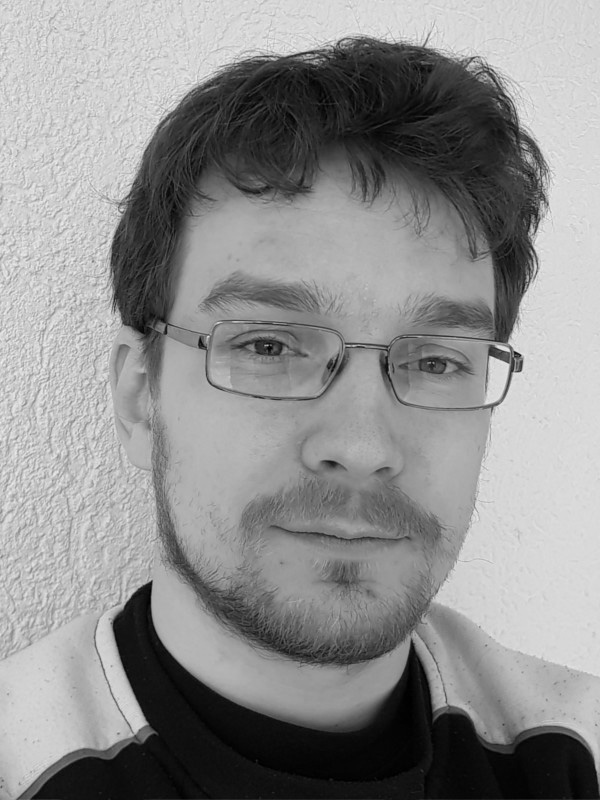}}]{Alexander Otte}
 worked at the mixedrealitylab at the University of Applied Sciences and Arts. He researched human-computer interaction in virtual and augmented reality with a focus on text entry. Specifically, he focused on building novel hardware prototypes and conduction human-subject experiments.
\end{IEEEbiography}

\begin{IEEEbiography}[{\includegraphics[width=1in,height=1.25in,clip,keepaspectratio]{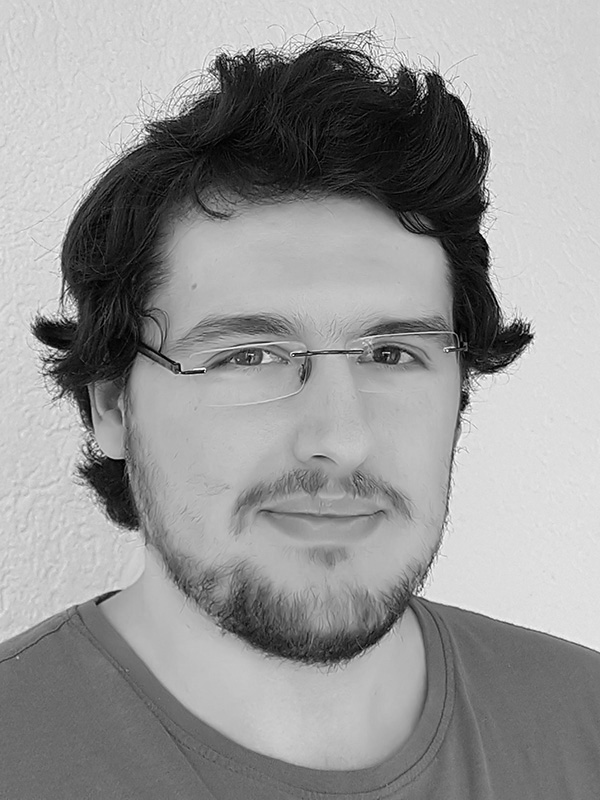}}]{Travis Gesslein}
 worked at the mixedrealitylab at the University of Applied Sciences and Arts. He researched interactive prototypes for augmented and virtual reality. Specifically, Travis worked on combining multiple input modalities for efficient cross-device interaction between head-mounted displays and touch devices such as smartphones and tablets.
\end{IEEEbiography}

\begin{IEEEbiography}[{\includegraphics[width=1in,height=1.25in,clip,keepaspectratio]{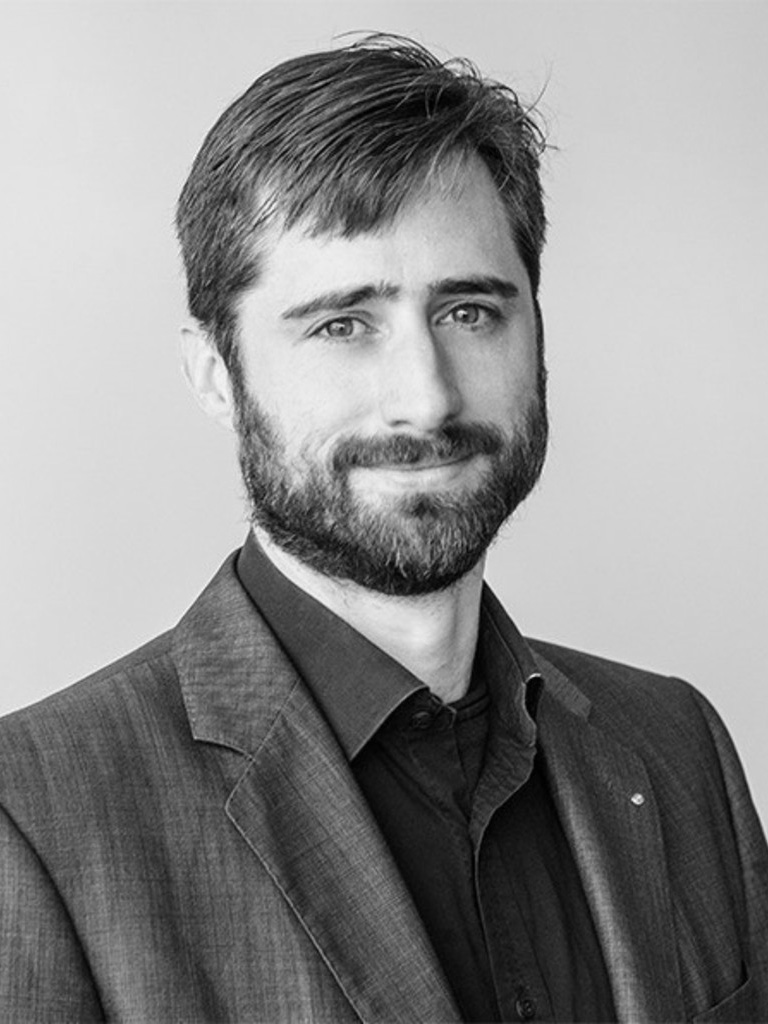}}]{Matthias Kranz}
is Professor of Embedded Systems at the Faculty for Informatics and Mathematics at Passau University. His research focuses on embedded systems and human-computer interaction.
\end{IEEEbiography}

\begin{IEEEbiography}[{\includegraphics[width=1in,height=1.25in,clip,keepaspectratio]{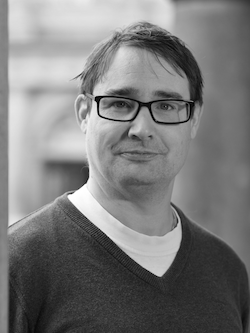}}]{Per Ola Kristensson}
is Professor of Interactive Systems Engineering in the Department of Engineering at the University of Cambridge and a Fellow of Trinity College, Cambridge. He leads the Intelligent Interactive Systems group, which belongs to the Engineering Design Centre. He is also a co-founder and co-director of the Centre for Human-Inspired Artificial Intelligence at the University of Cambridge.
\end{IEEEbiography}










\appendix 

\section{List of Abbreviations}

\nomenclature{AR}{Augmented Reality}
\nomenclature{CER}{Character Error Rate}
\nomenclature{CRT}{Cathode-ray Tube}
\nomenclature{FSS}{Flow-Short-Scale}
\nomenclature{HMD}{Head-mounted Display}
\nomenclature{HUD}{Heads-up Display}
\nomenclature{OST}{Optical See-through}
\nomenclature{RM-ANOVA}{Repeated Measures Analysis of Variance}
\nomenclature{SART}{Situational Awareness Rating Technique}
\nomenclature{SD}{Standard Deviation}
\nomenclature{SSQ}{Simulator Sickness Questionnaire}
\nomenclature{TLX}{Task Load Index}
\nomenclature{TOST}{Two One-sided Test}  
\nomenclature{VR}{Virtual Reality}

\printnomenclature

\end{document}